\documentclass[aps,prb,twocolumn,floatfix,notitlepage, superscriptaddress,longbibliography]{revtex4-2}
\usepackage[utf8]{inputenc}
\usepackage{amsfonts,amsmath,amssymb,graphicx,epstopdf,verbatim}
\usepackage{rotating}
\usepackage{epstopdf}
\usepackage{xcolor}
\usepackage{natbib}
\usepackage[colorlinks]{hyperref}
\definecolor{darkred}{rgb}{0.5,0,0}
\definecolor{darkgreen}{rgb}{0,0.5,0}
\definecolor{darkblue}{rgb}{0,0,0.5}
\hypersetup{colorlinks,
linkcolor=darkred,
filecolor=darkgreen,
urlcolor=darkblue,
citecolor=darkgreen}
\usepackage{braket}
\usepackage{overpic}

\newcommand{\Tr}{\mathrm{Tr}}

\newcommand{\calN}{\mathcal{N}}

\newcommand{\ud}{\mathrm{d}}
\newcommand{\nep}{\operatorname{e}}

\newcommand{\opcdag}[1]{{\hat{c}^{\dagger}}_{#1}}

\newcommand{\opc}[1]{{\hat{c}^{\phantom \dagger}}_{#1}}

\begin{document}

\title{Entanglement behavior and localization properties in monitored fermion systems}

\author{Giulia Piccitto}
\affiliation{Dipartimento di Matematica e Informatica, Università di Catania, Viale Andrea Doria 6, 95125, Catania, Italy}

\author{Giuliano Chiriac\`o}
\affiliation{Dipartimento di Fisica e Astronomia, Università di Catania, Via Santa Sofia 64, 95123, Catania, Italy}

\author{Davide Rossini}
\affiliation{Dipartimento di Fisica dell’Università di Pisa and INFN, Largo Pontecorvo 3, I-56127 Pisa, Italy}

\author{Angelo Russomanno}
\affiliation{Dipartimento di Fisica ``E. Pancini'', Universit\`a di Napoli Federico II, Complesso di Monte S. Angelo, via Cinthia, I-80126 Napoli, Italy}

\begin{abstract}
  We study the stationary bipartite entanglement in various integrable and nonintegrable models
  of monitored fermions evolving along quantum trajectories. We find that, for the integrable cases,
  the entanglement versus the system size is well fitted, over more than one order of magnitude, by a
  function interpolating between a linear and a power-law behavior. Up to the sizes we are able to reach,
  a logarithmic growth of the entanglement can also be captured by the same fit with a very small
  power-law exponent. For the nonintegrable cases, such as the staggered $t$-$V$ and the Sachdev-Ye-Kitaev
  (SYK) models, the numerical limitations prevent us from spanning different orders of magnitude in the system size.
  Here we fit the asymptotic entanglement versus the measurement strength with a generalized Lorentzian,
  finding a very good agreement, and then look at the scaling with the size of the fitting parameters. 
  We find two different behaviors: for the SYK we observe a linear increase with the system size,
  while for the $t$-$V$ model we see the emergence of traces of an entanglement crossover. In the latter models,
  we study the localization properties in the Hilbert space through the inverse participation ratio,
  finding an anomalous-delocalization behavior with no relation with the entanglement properties.
  Finally, we show that our function also fits very well the system-size dependence of the fermionic
  logarithmic negativity of a quadratic model in a two-leg ladder geometry, with stroboscopic projective measurements.
\end{abstract}

\maketitle

\section{Introduction}

Entanglement~\cite{Nielsen, RevModPhys.81.865} is a fundamental concept in quantum mechanics that has been
widely exploited to characterize many-body quantum systems. 
It can spotlight the presence of quantum phase transitions~\cite{Amico_RMP} and the existence of topological boundary
modes~\cite{Mondal_2022, PhysRevB.101.085136, PhysRevB.105.085106, micallo}, or can distinguish the unitary dynamics
of thermalizing, integrable, and many-body localized quantum
systems~\cite{Alba_2017, Alba_2018, Singh_2016, PhysRevB.102.144302, Yu_2016, Luitz_2016, De_Chiara_2006, PhysRevB.77.064426, PhysRevLett.109.017202, PhysRevB.101.064302}. 
Moreover, in short-range systems undergoing a unitary evolution, it attains at long times a value
that scales proportionally with the system size (volume law).

Recently, attention has shifted to understanding the effects of an external monitoring on the long-time steady-state behavior of the entanglement.
Intuitively, in the presence of projective measurements, the system collapses onto an eigenstate of the measurement operator
and the entanglement remains constant with the system size (area law, in one-dimensional systems). 
However this scenario may change when undergoing both a measurement processes and a Hamiltonian evolution. 
Indeed, the interplay between the entangling effect of the Hamiltonian dynamics and the disentangling role of measurements
leads to a variety of dynamical phases, characterized by peculiar entanglement behaviors.
This gives rise to the entanglement transitions that have been identified in a variety of models, spanning from quantum 
circuits~\cite{PhysRevX.10.041020, Li2018, Chan2019, Skinner2019, Szyniszewski2019, Vasseur2021, Bao2021, Nahum2020, Chen2020,Li2019, Jian2020, Li2021, Szyniszewski2020, Turkeshi2020, Lunt2021, Sierant2022_B, Nahum2021, Zabalo2020, Sierant2022_A, Chiriaco2023, Klocke2023,lirasolanilla2024},
to integrable or solvable~\cite{Nehra_2025, PhysRevB.109.214204, PhysRevResearch.6.043246, chahine2023entanglement, delmonte2024, Passarelli_2024, DeLuca2019, Nahum2020, Buchhold2021,Jian2022, Coppola2022, Fava2023, Poboiko2023, Jian2023, Merritt2023, Alberton2021, Turkeshi2021, Szyniszewski2022, Turkeshi2022, Piccitto2022, Piccitto2022e, Tirrito2022, Paviglianiti2023, Lang2020, Minato2022, Zerba2023, paviglianiti2023enhanced,chatterjee2024, gda_EPJB, Le_Gal_2024} and nonintegrable~\cite{Lunt2020,Rossini2020, Tang2020, Fuji2020, Sierant2021, Doggen2022, Altland2022, li2024monitored} Hamiltonian systems.

Focusing on free-fermion systems in the presence of local weak measurements, a crossover from a phase in which
the steady-state bipartite entanglement entropy (EE) grows logarithmically with the system size to a phase in which
it remains constant has been observed. The numerical evidence for a transition has been analytically confirmed in the case
of a $\mathbb{Z}_2$ symmetry~\cite{Fava2023} and challenged for a $U(1)$ symmetry~\cite{Poboiko2023},
suggesting that the logarithmic increase ceases for a size that is exponentially large in the inverse coupling,
thus being, in the latter case, just a finite-size crossover.
In the same context, when considering nonlocal weak measurements (e.g., power-law decaying measurement operators)
or particular lattice geometries, one numerically observes transitions between three distinct situations: A volume-law,
an intermediate subvolume-law, and an area-law entanglement phase~\cite{Russomanno2023_longrange, Tsitsishvili_2024}.
This is similar to what happens when the dynamics is only induced by random measurements of nonlocal
strings~\cite{Ippoliti2021, Sriram2022, Piccitto2023}.

Here we look at these phenomena from a different perspective. We start considering some integrable
monitored fermionic systems (including the ones mentioned above)
coupled to the environment through a quantum-state-diffusion monitoring process. More specifically, we study (i) the tight-binding chain with onsite dephasing, (ii) the Kitaev chain with onsite dephasing,
and (iii) the Kitaev chain with long-range dissipators.
For these models the EE has been already investigated in the literature.
Here we consider the problem from a slightly different perspective and find that the steady-state value of the bipartite EE
depends on the system size $L$ in a way that is well fitted, for all the sizes we study,
by the function
\begin{equation}
  f(L) = \frac{A \, L}{1 + C \, L^b}, \qquad (A,\, C,\, b \geq 0).
  \label{eq:fit_function}
\end{equation}

The rationale behind this fitting function is the following.
For small system sizes, one expects a linear increase of the entanglement entropy with $L$.
The reason is that, {in the absence of dissipation, the entanglement entropy increases linearly in time,
  due to the local-Hamiltonian ballistic propagation of correlations~\cite{Alba_2018,lightcone2}, and this propagation
  is stopped by the environment measurements, on average after
  a characteristic time scale. This is especially true in the integrable systems we focus on, 
  due to the ballistic propagation of quasiparticles~\cite{calabrese_notes}. In the presence of dissipation, this propagation stops on average after a characteristic time scale, due to the environment measurements~\cite{DeLuca2019}. So, we expect
  to see a linear increase of the steady-state entanglement entropy with the size, if $L$ is smaller than $v_0 t^*$, where $t^*$ is the characteristic time after which the propagation stops and $v_0$ is the quasiparticle propagation velocity. In fact, for $L<v_0 t^*$, quasiparticles (and correlations they bring with them) have the time to saturate the full chain before their propagation is arrested, thus the long-time entanglement entropy is linear in $L$.}
In agreement with that, a linear increase in $L$ for small system sizes is exactly provided
by Eq.~\eqref{eq:fit_function} in the limit $L \ll C^{-1/b}$.
On the opposite hand, for large system sizes ($L \gg C^{-1/b}$), Eq.~\eqref{eq:fit_function} gives rise
to a power-law behavior $f(L) \sim \frac{A}{C}L^{1-b}$. We should stress that the fit is reliable as long as
the numerical points encompass a range with a maximum $L_{\rm max}$ larger or equal than the length scale 
\begin{equation}\label{l0:eqn}
  L_0\equiv C^{-1/b}
\end{equation}
separating the short-range linear from the long-range power-law behavior.
For all the fits we perform with Eq.~\eqref{eq:fit_function}, we check that this condition is verified.
In the long-range power-law regime we find different behaviors (depending on the value of $b$) for different models.

In particular, in the tight-binding chain with onsite dephasing it is well known that the system asymptotically in the system size always obeys an area-law behavior~\cite{DeLuca2019,Poboiko2023}.
In this case therefore we fit the steady-state EE with the function Eq.~\eqref{eq:fit_function} fixing $b=1$ so that the large-system-size behavior is an area law. We find a very good agreement
between the fitting function and the numerical point, fully confirming the theoretical predictions. We numerically find that $L_0$ scales as a power law in $\gamma$, not so far from the $\gamma^{-1}$ predictions of Ref.~\cite{DeLuca2019}.

The Kitaev chain with and long-range dissipators is peculiar as well. Here we previously found~\cite{Russomanno2023_longrange} a volume-law regime, an area-law regime, and an intermediate subvolume regime
that impossible to fit logarithmically. We find that the fit with Eq.~\eqref{eq:fit_function} works fine in the three regimes, providing an asymptotic linear increase ($b=0$)
in the volume law regime, an asymptotic finite value ($b=1$) in the area law regime, and an asymptotic power-law behavior ($0<b<1$) in the intermediate regime. In the
intermediate regime $b$ (and the power-law exponent $1-b$) displays a plateau. 
A similar power-law behavior, called ``superdiffusive behavior'', has already been observed in a model~\cite{poboiko2025measurementinducedlevyflightsquantum} where symmetries allow quasiparticles to propagate long enough
to enhance the correlations and increase the steady-state entanglement. In our model correlations are instead physically created by the long-range measured operators 
decaying as a power law. Moreover, we find that $L_0$ diverges when $\alpha$ tends to 1 from above. Indeed, for $\alpha<1$, the system displays a volume-law behavior of entanglement and quite consistently the scale $L_0$ separating volume from power-law behavior tends to infinity.

We point out that, at the present stage, our findings are only based on a purely numerical analysis,
therefore Eq.~\eqref{eq:fit_function}
should not be exploited to classify the various entanglement phases.
Despite this, as we shall see below, the remarkably good agreement of such fitting formula with the numerical
data (even in the small-size range, differently from other proposed scaling {\em Ans{\"a}tze})
suggests that Eq.~\eqref{eq:fit_function} may motivate future investigations
on the entanglement dynamics in monitored systems.

Then we move on and extend our analysis to two nonintegrable models: (iv) the $t$-$V$ staggered model
and (v) the Sachdev-Ye-Kitaev (SYK) model~\cite{Sachdev_1993, Kitaev_2015},
which have been recently considered in the context of entanglement transitions~\cite{xing2023interactions, PhysRevLett.127.140601}.
Although their nonintegrability prevents us from accessing large sizes, we have worked out some scaling by fitting
with a generalized Lorentzian the asymptotic entanglement at fixed $L$ versus the coupling $\gamma$ to the environment.
Then, we look at the scaling of the fitting parameter with the system sizes, finding that 
the $L$-dependence of Eq.~\eqref{eq:fit_function} is recovered.
Collecting all the results, we obtain that for the fully chaotic SYK model, the EE linearly increases with the system size
at any measurement strength. 
Conversely, in the $t$-$V$ staggered model, we find hints of an entanglement crossover in $\gamma$.

We also address the localization properties of these models. Analyses of localization/delocalization
properties in some integrable monitored systems already exist~\cite{Marcin,gda_EPJB,chahine2023entanglement};
here we extend this analysis to nonintegrable cases. Namely, we study the time- and realization-averaged
logarithm of the inverse participation ratio (IPR) in the Hilbert space and find that this quantity scales linearly
with the logarithm of the dimension of the Hilbert space, with a slope that depends on $\gamma$.
Its value corresponds neither to perfect delocalization nor to perfect
localization, but rather to an anomalous delocalization, akin to a multifractal behavior. 
This qualitative picture holds also when moving to the integrable limit and is independent of the entanglement
behavior, suggesting that localization properties are not related to the entanglement transitions.

To witness the applicability of our procedure in a wider context, in the third part we show that it also applies
to the fermionic logarithmic negativity (FLN). To do so, we focus on a noninteracting fermionic model on a two-leg ladder,
undergoing projective measurements at discrete times~\cite{Tsitsishvili_2024, muzzi2024}.
As for the EE, we find that the asymptotic FLN versus the system size is well described by Eq.~\eqref{eq:fit_function},
thus we are able to recognize traces of the different dynamical regimes of the entanglement through the behavior of the FLN.

The paper is organized as follows. In Sec.~\ref{moni:sec} we briefly recall the Lindblad description of monitored
fermionic systems, focusing in particular on the quantum state diffusion protocol. In Sec.~\ref{entro:sec} we define
the asymptotic bipartite EE and describe the proposed function to characterize its behavior.
Then we present our results for integrable (Sec.~\ref{models:sec}) and for nonintegrable models (Sec.~\ref{models_ni:sec}).
Finally, in Sec.~\ref{Sec:Giuliano_model} we focus on the FLN in a ladder fermionic model.
Our conclusions are drawn in Sec.~\ref{conc:sec}.  In the Appendix we provide further details
on the stability of our fit (App.~\ref{App:stability}), on the time traces of the
trajectory-averaged EE (App.~\ref{ttraces:sec}), and on how to compute the FLN
for the monitored noninteracting fermionic ladder (App.~\ref{App:Evolution}).

\section{Monitored fermionic systems}
\label{moni:sec}

We consider systems of spinless fermions on a lattice with $L$ sites, described by Hamiltonians which
can be generically cast as the sum of a quadratic and (possibly) a quartic term
$\hat H = \hat H^{(2)} + \hat H^{(4)}$, where we define
\begin{subequations}
  \label{eq:Hamiltonians}
  \begin{align}
    \hat H^{(2)} = &\sum_{i,j=1}^L \big( D_{ij} \, \hat c_i^\dagger \hat c_j + O_{ij} \, \hat c^\dagger_i \hat c^\dagger_j + \text{h.c.} \big) \,, \\
    \hat H^{(4)} = &\sum_{i,j,k,l=1}^L \big( J_{ij,kl} \, \hat c_i^\dagger \hat c^\dagger_j \hat c_k \hat c_l + \text{h.c.} \big) \,. 
  \end{align}
\end{subequations}
The operators $\hat c^{(\dagger)}_j$ annihilate (create) a fermion on the $j$th site and obey the canonical
anticommutation relations
\begin{equation}
  \{\hat c_i, \hat c_j^\dagger\} = \delta_{ij}, \quad \{\hat c_i, \hat c_j\} = 0 \,.
\end{equation}
To ensure Hermiticity, the complex coupling constants in Eqs.~\eqref{eq:Hamiltonians}
must respect the following constraints:
\begin{subequations}
  \begin{align}
    & D_{ij} = D_{ji}^*, \quad O_{ij} = -O_{ji} \,, \\
    & J_{ij,kl} = - J_{ji,kl} = - J_{ij,lk} = J^*_{lk,ij} \,.
  \end{align}
\end{subequations}
The $\hat H^{(2)}$ term is quadratic in the creation/annihilation operators $\{ \hat c^{(\dagger)}_j \}$ and is integrable,
while the $\hat H^{(4)}$ term introduces correlations between fermions and breaks integrability.
In what follows, we consider four different integrable Hamiltonians [with $\hat H^{(4)} = 0$]
and two nonintegrable ones [with $\hat H^{(4)} \neq 0$].
Details on the various models are provided in Secs.~\ref{models:sec}-\ref{Sec:Giuliano_model}
and~\ref{models_ni:sec}, respectively.

We are interested in describing the dynamics in the presence of weak measurements of some Hermitian operator $\hat m_j$.
As is known~\cite{Daley2014, Plenio, fazio2024manybodyopenquantumsystems}, a single realization of the measurement sequence
can be described by the stochastic evolution of a pure state $\ket{\psi(t)}$ (namely, a quantum trajectory).
On average, the system is described by a density matrix $\rho_t = \overline{\ket{\psi(t)}\bra{\psi(t)}}$
(the overline indicates ensemble averaging over many trajectories) obeying a Lindblad master equation
\begin{equation}
  \partial_t\rho_t \!= -i [\hat{H}, \rho_t]
  + \gamma \!\sum_j \!\Big( \hat{m}_j \, \rho_t \, \hat{m}_j - \tfrac{1}{2} \{ \hat{m}_j^2,\rho_t \} \Big)\,,
  \label{eq:lind}
\end{equation}
where $\gamma$ represents the system-environment coupling. Hereafter we use units of $\hbar = 1$.
We remind that there are many choices of stochastic-dynamics protocols, also known as unravelings,
that provide the same average state $\rho_t$. Different stochastic evolutions mimic different measurement protocols.

Except for Sec.~\ref{Sec:Giuliano_model} (whose details are given later), the results in Sec.~\ref{models:sec}
and in Sec.~\ref{models_ni:sec} are obtained by implementing a composite dynamics given by
(i) an Hamiltonian evolution following a quantum quench
and by (ii) a process of continuous measurement of the operator $\hat m_j$ (with $j = 1, \dots, L$).
The dynamics, known as quantum state diffusion, along each trajectory can be obtained by integrating
the stochastic equation~\cite{Daley2014, Plenio, fazio2024manybodyopenquantumsystems}
\begin{eqnarray}
  d \ket{\psi(t)} & = & - \Big[ i \hat H + \sum_j \frac{\gamma}{2} \big( \hat m_j - \braket{m_j}_t \big)^2 \Big] dt \ket{\psi(t)} \nonumber \\	
    && + \Big[ \sum_j \sqrt{\gamma} \, \big( \hat m_j - \braket{\hat m_j}_t \big) \, dW_t^j \Big] \ket{\psi(t)},
  \label{schro:eqn}
\end{eqnarray}
where $\braket{\, \cdot \,}_t \equiv \braket{\psi(t)|\cdot|\psi(t)}$, while $W_t^j$ are independent
Wiener processes (for $j=1,\ldots,L$). 
The state $\ket{\psi(t)}$ along each trajectory appearing in Eq.~\eqref{schro:eqn} is called the unraveled state.
We can discretize the evolution time with steps of length $\delta t$ and Trotterize the evolution.
In what follows we consider measurement operators having the property 
\begin{equation}
  \label{bono:eqn}
  \hat m_j^2 = p_j+q_j\hat{m}_j\,, \qquad \text{with} \quad p_j,\,q_j \in \mathbb{R}\,.
\end{equation}
Under this assumption, up to $o(\delta t)$ terms, we get the expression~\cite{DeLuca2019}
\begin{equation}
  \label{step:eqn}
  \ket{\psi(t+\delta t)} \! \approx \mathcal{C} \  e^{\sum_j \!\! \big[ \delta W^j_t + (2 \braket{\hat m_j}_t -q_j)\gamma \delta t\big] \hat m_j} 
  e^{-i\hat H \delta t} \! \ket{\psi(t)}\!\,,
\end{equation}
where the constant $\mathcal{C}$ normalizes the evolved state.  
The $\delta W^j_t$ are zero-mean Gaussian random variables with
$\braket{\delta W_l(t) \, \delta W_j(t')}_t = \gamma \, \delta t \, \delta_{lj} \, \delta_{tt'}$.
The Lindblad master equation~\eqref{eq:lind} can be recovered by averaging over the quantum trajectories
and performing the limit $\delta t\to 0$.

Coming to the choice of the initial state, in all simulations we start from the staggered N\'eel state
\begin{equation}
  \label{init:eqn}
  \ket{\psi(0)} = \prod_{j=1}^{L/2} \hat c_{2j}^\dagger \ket{\Omega},
\end{equation}
where $\ket{\Omega}$ is the vacuum state for the $\hat c_j$-fermions.
Note that, in general, this is not the ground state of the Hamiltonian $\hat H$ inducing the unitary part
of the dynamics, so in this sense we are applying a quantum quench.

\section{Asymptotic averaged bipartite entanglement entropy}
\label{entro:sec}

To access the asymptotic averaged entanglement, we consider a partition of the global system into two
subsystems $A$ and $B$ of length $\ell$ and $L - \ell$, respectively. We can thus compute
the von Neumann entropy of one subsystem~\cite{Nielsen},
\begin{equation}
  S_{\ell}(t) = - \Tr \big[ \rho_A(t) \ln \rho_A(t) \big] \,, 
\end{equation}
being $\rho_A(t) = \Tr_B \big[ \ket{\psi(t)}\bra{\psi(t)} \big]$ the reduced density matrix of subsystem $A$.
Provided the global system is in a pure state [$\ket{\psi(t)}$, in this case], the quantity $S_\ell(t)$
is a good measure of the entanglement between $A$ and $B$ and is usually referred to as the bipartite EE.
Then, we average over many quantum trajectories
\begin{equation}
	\overline{S_{\ell}(t)} = - \overline{\Tr \big[ \rho_A(t) \ln \rho_A(t) \big] }.
\end{equation}
Notice that this operation is different from evaluating the von Neumann entropy over the average state
$\rho_t = \overline{\ket{\psi(t)}\bra{\psi(t)}}$ which, besides that, would also not be
a proper measure of the entanglement. Finally, we fix $\ell$ to be a fixed fraction of $L$
(in particular we consider either $\ell = L/2$ or $\ell = L/4$) and estimate the asymptotic
long-time value of $\overline{S_{\ell}(t)}$ by performing a suitable time average:
\begin{equation}
  \label{selle:eqn}
  \overline{S}_\ell = \frac{1}{t_f-t_0}\int_{t_0}^{t_f} \overline{S_\ell(t)} \, \ud t\,.
\end{equation}
Here $[t_0, t_f]$ is an appropriate time window in which the behavior of $\overline{S_\ell(t)}$ has attained a steady-state value.
We perform the average over quantum trajectories numerically, over a finite number of realizations
that we fix as $N_{\rm r} = 48$, when not differently specified.
The error bars for our data are evaluated as the standard error (root-mean square deviation divided by $\sqrt{N_{\rm r}}$).

We aim at studying the dependence on the system size of the asymptotic averaged EE in Eq.~\eqref{selle:eqn}.
Apart from few notable exceptions~\cite{Poboiko2023, Fava2023, PhysRevResearch.6.043246}, analytical models allowing
for a {\it a priori} determination of the scaling regime are lacking, and then one must rely on numerical analysis
that is usually limited to small $L$. 
Here we propose to use the function in Eq.~\eqref{eq:fit_function} for fitting the behavior of $\overline{S}_\ell$ versus $L$,
determining the parameters $A$, $C$, and $b$ by a fit of the numerical data.
The function interpolates between a linear and a power-law dependence of $\overline{S}_\ell$ with $L$, for increasing the size.
In particular, for $L \gg 1$ we have
\begin{equation}
  \overline{S}_\ell \sim \frac{A}{C}L^{1-b}.
  \label{eq:fit_largeL}
\end{equation}
Therefore, the dynamical regime is encrypted in the behavior of the parameter $b$, in the following way:
\begin{equation}
  \label{regimes:eqn}
  \begin{aligned}
    b = 0,     & \qquad \text{for a volume-law,}\\
    0 < b < 1, & \qquad \text{for a subvolume-law,}\\
    b \ge 1,   & \qquad \text{for an area-law}.
  \end{aligned}
\end{equation}

In what follows, we fit $\overline{S}_\ell$ versus $L$ for different fermionic models with Eq.~\eqref{eq:fit_function}.
The most interesting result is that this function seems to fit our numerics very well, independently of the considered model.
In particular, in the next section we specialize to integrable models, while in Sec.~\ref{models_ni:sec}
we focus on nonintegrable models.
For the latter, due to the small attainable system sizes ($L \lesssim 20$), it is more convenient
to fit $\overline{S}_\ell$ versus the coupling $\gamma$ with the environment. We choose a generalized Lorentzian function
and, in the end, we find that the parameters of the fit scale with the system size, in such a way that
the form in Eq.~\eqref{eq:fit_function} is recovered (details are provided in Sec.~\ref{nonintegrable:sec}). 
Finally, in Sec.~\ref{Sec:Giuliano_model}, we switch to an free-fermion model with a more complicated geometry
and a stroboscopic evolution, for which we consider the entanglement between two portions of a part
of the whole system: as in that case the relevant part of the system is described by a mixed state,
we resort to a proper entanglement monotone such as the FLN.
To keep the presentation more accessible, we postpone all the required definitions to that section.

\section{Integrable models}
\label{models:sec}

We first focus on integrable fermionic models, whose dynamics can be reliably accessed
up to quite large system sizes ($L \lesssim 10^{3}$), thanks to the Gaussianity property,
and the fit of the asymptotic averaged EE with Eq.~\eqref{eq:fit_function} is meaningful. 
In the following, we consider three models on a one dimensional lattice, whose Hamiltonian is of the type $\hat H = \hat H^{(2)}$.
Namely, the tight binding chain with local dephasing (Sec.~\ref{U1:par}), the Kitaev chain (Sec.~\ref{Z2:par})
again with local dephasing, and the Kitaev chain with long-range dissipators (Sec.~\ref{Z2-lr:par}),
and study the entanglement behavior for each of these situations.

\subsection{Tight-binding chain with onsite dephasing}
\label{U1:par}

We start with a simple tight-binding chain, described by a nearest-neighbor hopping Hamiltonian
and subject to local (onsite) dephasing~\cite{DeLuca2019}:
\begin{subequations}
  \label{U1_int:eqn}
  \begin{align}
    \hat H_{\rm t\mbox{-}b} & = -\frac{J}{2} \sum_{j=1}^L \big( \hat{c}_j^\dagger\hat{c}_{j+1} + {\rm h.c.} \big) \,,
    \label{U1_int_Ham} \\
    \hat{m}_j & = \hat n_j, \qquad \text{for} \quad j = 1,\,\ldots,\,L\,,
    \label{dephas}
  \end{align}
\end{subequations}
where $J$ denotes the hopping strength and $\hat n_j = \hat c^\dagger_j \hat c_j$ is the
onsite fermion number operator. 
Here and in the other considered one-dimensional models, we adopt periodic boundary conditions
by assuming $\hat c^{(\dagger)}_{L+1} \equiv \hat c^{(\dagger)}_1$.
With reference to Eq.~\eqref{bono:eqn}, we have $p_j = 0$ and $q_j = 1$.
This system possesses a $U(1)$ symmetry, corresponding to the conservation of the total number of fermions,
$\hat{N} = \sum_j \hat n_j$.

In this case, the unraveled state $\ket{\psi(t)}$ can be always cast in a Slater determinant form~\cite{DeLuca2019, gda_EPJB}
\begin{equation}
  \ket{\psi(t)} = \prod_{k=1}^N \bigg[ \, \sum_{j=1}^L \big[ U_t \big]_{jk} \, \hat c_j^\dagger \bigg] \ket{\Omega}\,,
  \label{eq:Slater}
\end{equation}
so that one ends up with the study of the dynamics of the $L \times N$ matrix $U_t$,
a problem which scales polynomially (and not exponentially) with $L$.
[Starting from the Néel state~\eqref{init:eqn}, we have $N=L/2$, so that $U_t$ is a $L\times L/2$ matrix.]
As a consequence, quite large system sizes can be reached numerically, up to some hundreds.

\begin{figure}[!t]
  \includegraphics[width=8cm]{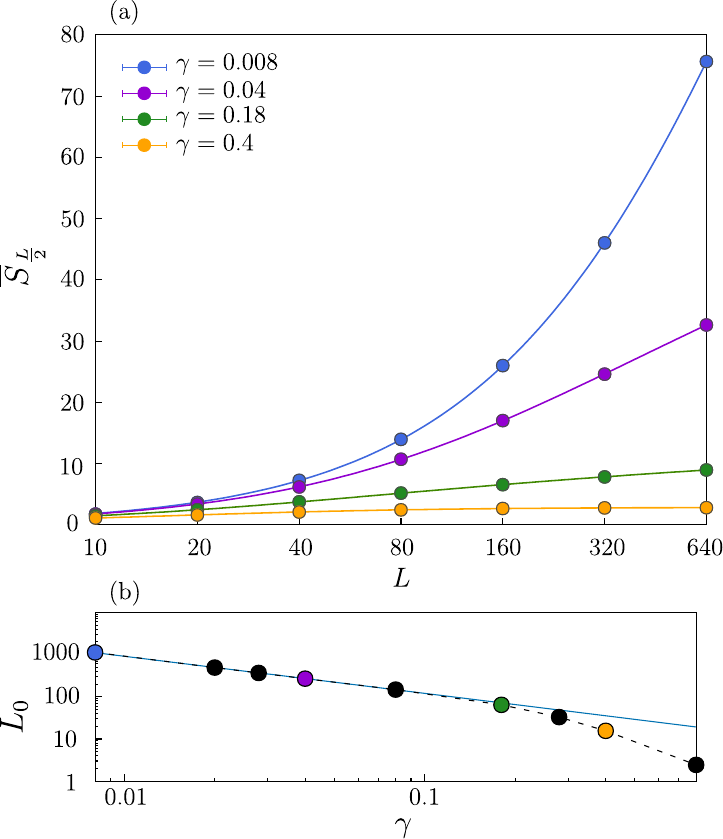}
  \caption{The asymptotic averaged EE for the model in Eqs.~\eqref{U1_int:eqn}.
    (a) Some examples with the behavior of $\overline{S}_{L/2}$ versus $L$ (circles) for different values
    of $\gamma$, together with the corresponding fits with Eq.~\eqref{eq:fit_function} (continuous lines)
    where we fix the parameter $b$ as $b=1$.
    (b) The length scale $L_0=1/C$ ---with $C$ obtained by fitting the curves in panel (a)
    with Eq.~\eqref{eq:fit_function}---
    versus $\gamma$ in a double-logarithmic plot. The straight line results from the fit of the data
    for $\gamma < 0.1$ and corresponds to a power law of the form $L_0\sim \gamma^{-0.88}$.
    We simulate the time evolution until $t_f = 6 \times 10^6$, with a step $\delta t = 0.01$.
    Here and in the next figures, we work in units of $J=1$.}
  \label{fitI:fig}
\end{figure}

For the model in Eqs.~\eqref{U1_int:eqn}, the authors of Ref.~\cite{DeLuca2019} showed the existence of an area-law phase
for the asymptotic EE. More recently, the existence of a transition from area- to logarithm-law
has been first claimed~\cite{Alberton2021, Ladewig2022} and then challenged. 
In fact, through the replica trick within a Keldysh path-integral formalism, it has been suggested that
only the area-law phase exists, while the logarithm-law phase should be just a finite-size crossover,
due to the exponential growth of a localization length with the inverse measurement strength~\cite{Poboiko2023}.
For this reason, we fix $b=1$ in the fitting function~\eqref{eq:fit_function},
so that asymptotically for large system sizes one gets an area law.
     
Figure~\ref{fitI:fig}(a) displays numerical results for $\overline{S}_{L/2}$ versus $L$ (circles), for some values
of the measurement strength $\gamma$, and the corresponding fit of Eq.~\eqref{eq:fit_function} (continuous lines). 
We observe an accurate agreement between the two. This finding confirms the asymptotic area-law prediction,
considering that the fit is meaningful because the maximum of the fitting range is comparable
to the scale $L_0 = 1/C$ separating the short-range from the long-range behavior.
This is evaluated using the value of $C$ extracted from the fit and is plotted versus $\gamma$
in a double-logarithmic plot [see Fig.~\ref{fitI:fig}(b)]. Notice that, for $\gamma < 0.1$, the data tend
to behave as a power law (straight line in the log plot). Applying a least-square linear fit we get
$L_0\simeq \gamma^{-0.88}$. This result is not very far from the behavior $L_0 \simeq 1/\gamma$ predicted
in Ref.~\cite{DeLuca2019}.

\subsection{Kitaev chain with onsite dephasing}
\label{Z2:par}
We now discuss the one-dimensional Kitaev model~\cite{Kitaev_2001} with
the same local dephasing~\cite{Piccitto2022, Turkeshi2021}:
\begin{subequations}
  \label{Z2_int:eqn}
  \begin{align}
    \hat H_{\rm K} & = - \sum_{j=1}^L\left[{J} \big( \hat{c}_j^\dagger\hat{c}_{j+1} +
      \hat{c}_j^\dagger\hat{c}_{j+1}^\dagger + {\rm h.c.} \big) + 2 h \hat n_j \right], \label{eq:Ham_Z2_int} \\
    \hat{m}_j & = \hat n_j , \qquad \text{for} \quad j = 1,\,\ldots,\,L\,,
  \end{align}
\end{subequations}
where $J$ is the nearest-neighbor coupling and $2h$ is a local chemical potential. 
This system has a $\mathbb{Z}_2$ symmetry, since the parity $\hat P = \prod_j \hat n_j$
of the fermion number is conserved (the number of particles $\hat N$ itself is not conserved,
due to the presence of the pairing terms $\hat{c}_j^\dagger \hat{c}_{j+1}^\dagger$).

The form of the unraveled state $\ket{\psi(t)}$ is slightly different from the Slater determinant~\eqref{eq:Slater}
and can be cast in the following Gaussian shape~\cite{Mbeng_SciPostPhysLectNotes24}:
\begin{equation}
  \label{statog:eqn}
  \ket{\psi(t)} = {\calN}_t \, \exp{\bigg[ \frac{1}{2} \sum_{j_1,j_2=1}^L \big[ { Z}_t \big]_{j_1 j_2} \opcdag{j_1} \opcdag{j_2}\bigg] }
  \, |0\rangle\,,
\end{equation}
where ${\cal N}_t$ is a normalization prefactor and ${ Z}_t$ is an antisymmetric $L\times L$ matrix
that can be written as ${ Z}_t= -[{ U}_t^\dagger]^{-1} \, { V}_t^\dagger$.
The $U_t$ and $V_t$ can be cast as the submatrices of a Bogoliubov rotation allowing to construct the fermionic operators
that annihilate the unraveled state~\eqref{statog:eqn} and obey linear differential
equations~\cite{Mbeng_SciPostPhysLectNotes24, Russomanno2023_longrange}.
The interpretation as a Bogoliubov rotation is valid if ${U}_t$ and ${V}_t$ obey a unitarity condition,
a constraint that can be restored (keeping ${Z}_t$ unchanged) by using a QR decomposition~\cite{Russomanno2023_longrange}.
One can therefore restrict to study the dynamics of the two $L \times L$ matrices ${U}_t$ and ${V}_t$,
keeping a polynomial scaling of the problem complexity and thus allowing the numerics to reach systems
with a few hundreds of sites. 

The monitored dynamics of the model in Eqs.~\eqref{Z2_int:eqn}
has been widely studied from a numerical point of view~\cite{Turkeshi2021, Turkeshi2022, Piccitto2022}, always supporting
a transition from a logarithm-law to an area-law regime, depending both on the measurement strength and on the parameter $h$
in the Hamiltonian~\eqref{eq:Ham_Z2_int}. This transition has been also analytically proved by exploiting
an approach based on a nonlinear sigma model~\cite{Fava2023}.

\begin{figure}[!t]
  \includegraphics[width=8cm]{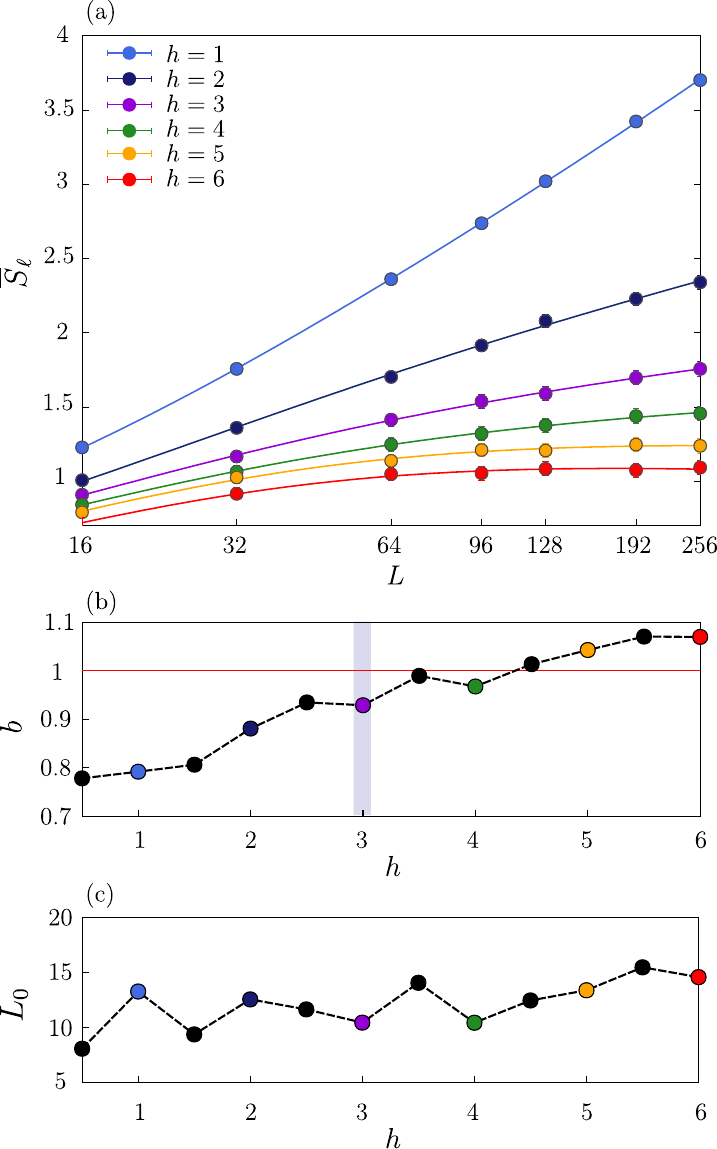}
  \caption{The EE for the model in Eqs.~\eqref{Z2_int:eqn}.
    (a) $\overline{S}_{L/4}$ versus $L$ (circles) for different values of $h$ and fixed $\gamma=1.5$,
    together with the corresponding fits with Eq.~\eqref{eq:fit_function} (continuous lines).
    (b) The fitting parameter $b$ ---obtained by fitting the curves in panel (a) with Eq.~\eqref{eq:fit_function}---
    versus $h$. (c) The length scale $L_0$ [see Eq.~\eqref{l0:eqn}] versus $h$ obtained with the numerical fit.
    We evolve up to $t_f = 60$ with a step $\delta t = 0.05$ and take $N_{\rm r}=100$.}
  \label{fig:ising}
\end{figure}

In Fig.~\ref{fig:ising}(a) we show our numerical results for $\overline{S}_{L/2}$ versus $L$ (circles), for some
values of $h$ and fixed $\gamma$, and the corresponding fit obtained using Eq.~\eqref{eq:fit_function} (lines).
Even in this case we observe a nice agreement between the numerical data and the fitting function,
over all the considered range of sizes $16 \leq L \leq 192$. In Fig.~\ref{fig:ising}(c) we plot $L_0$ versus $h$ and
see that it is always significantly smaller than the maximum of the range of sizes where we apply the fit 
[$L_{\rm max} = 256$ as shown in Fig.~\ref{fig:ising}(a)], thus confirming the reliability of our fit.
Looking in more detail at the fit exponent $b$ as a function of $h$ [Fig.~\ref{fig:ising}(b)],
we observe a monotonically increasing behavior, contrary to that for the tight-binding model reported
in Fig.~\ref{fitI:fig}. We also see that $b$ now gets significantly smaller than one, for small values of $h$.
So we can more confidently state that, in this case, there is a large-$h$ regime where the EE displays
and area-law behavior and a small-$h$ regime where the EE is likely scaling with the system size
in a sub-volume way. 
Unfortunately, it is difficult to mark a precise crossover point: in the region where $b$
is slightly smaller than one, the same issues occurring for the model of Sec.~\ref{U1:par} emerge.
In particular, with the available system sizes, it is impossible to distinguish between
an area-law and a logarithm-law behavior.
Curiously, the threshold value $h \approx 3$, conjectured to be the crossover point (for $\gamma=1.5$)
on the basis of an alternative fit of the numerical data up to $L=256$ performed in Ref.~\cite{Piccitto2022},
is compatible with the analysis reported in Fig.~\ref{fig:ising}(b).
However we stress that, despite the procedure in Ref.~\cite{Piccitto2022} was rather sensitive to
finite-size effects, the one outlined here seems to us more appropriate and robust in this sense
(see Appendix~\ref{App:stability} for details on the numerical stability of the fits).
Here we have only discussed the case $\gamma=1.5$, although we checked that analogous considerations apply
for other values of the system-bath coupling, leading to the same qualitative conclusions (not shown).

\subsection{Kitaev chain with long-range dissipators}
\label{Z2-lr:par}

A nonlocal-measurement extension of the previous case can be obtained by keeping the same Hamiltonian $\hat H_{\rm K}$
as in Eq.~\eqref{eq:Ham_Z2_int}, but using long-range Lindblad operators which decay as a power-law with the distance.
More specifically, the jump operators are given by~\cite{Russomanno2023_longrange}:
\begin{equation}
\begin{aligned}
  \label{Z2_int_lr:eqn}
    &\hat{m}_i  = \sum_{j=1}^L f_{ij} \big( \opc{i}-\opcdag{i} \big) \big( \opc{j}+\opcdag{j} \big), \\
    &f_{ij} = \frac{1}{N(\alpha)}\frac{1}{(1+D_{ij})^\alpha}, \quad \text{for} \quad i,j = 1,\,\ldots,\,L\,,
  \end{aligned}
\end{equation}
with $\alpha \geq 0$, and $N(\alpha) \equiv (N-1)^{-1} \sum_{i, j} (1+D_{ij})^{-\alpha}$ being the Kac normalization factor.
Here $D_{ij}$ is the distance between the $i$th and the $j$th site. Since we are considering periodic boundary conditions,
we assume $D_{ij} = \min(|i-j|, N-|i-j|)$. With reference to Eq.~\eqref{bono:eqn} we have $q_j = 0$ and $p_j = \sum_l f_{jl}^2$.

Also in this model the $\mathbb{Z}_2$ symmetry associated to the parity is preserved and, due to the particular
structure of the measurement operators, the quantum-state-diffusion dynamics preserves the Gaussianity
of the unraveled state $\ket{\psi(t)}$, that can be cast in the form Eq.~\eqref{statog:eqn}.
Previous numerical investigations of the dynamics of this model already showed the emergence of three
parameter regions where the EE behaves distinctly, ranging from volume-law, to area-law,
as well as to subvolume-law scaling with the system size~\cite{Russomanno2023_longrange}. 
The subvolume scaling occurs in an intermediate region between
the area-law and the volume-law ones: this corresponds to a steady-state EE exhibiting
a less-than-linear growth that, differently from more common cases, is faster than logarithmic.

\begin{figure}[!t]
  \includegraphics[width=8cm]{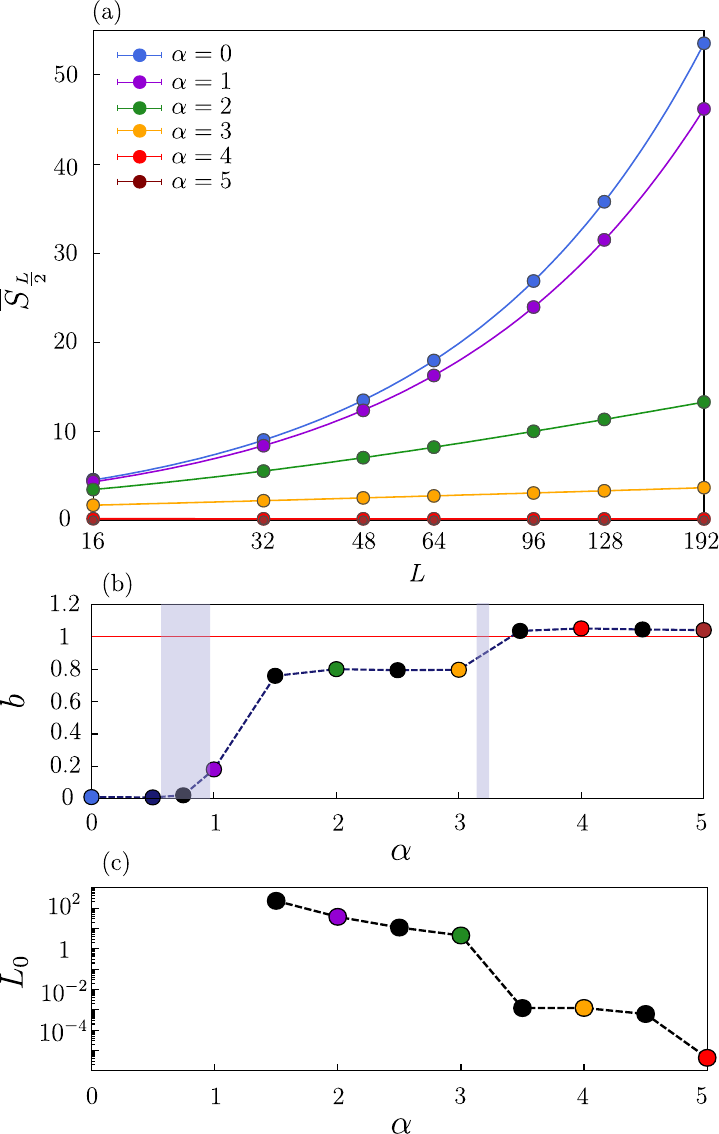}
  \caption{The EE for the model in Eqs.~\eqref{eq:Ham_Z2_int} and~\eqref{Z2_int_lr:eqn}.
    (a) $\overline{S}_{L/2}$ versus $L$ (circles) for different values of $\alpha$ and fixed $\gamma = 0.1$, $h = 0.5$,
    together with the corresponding fits with Eq.~\eqref{eq:fit_function} (continuous lines).
    (b) The fitting parameter $b$ ---obtained by fitting the curves in panel (a) with Eq.~\eqref{eq:fit_function}---
    versus $\alpha$.
    (c) The length scale $L_0$ [see Eq.~\eqref{l0:eqn}] versus $\alpha$ obtained with the numerical fit.
    We simulate the time evolution until $t_f = 3 \times 10^3$,
    with a step $\delta t = 0.005$.
    The shaded areas locate the two crossover regions found in Ref.~\cite{Russomanno2023_longrange}
    for the same set of parameters used here.}
  \label{long:fig}
\end{figure}

All these regimes can be recognized by fitting the asymptotic averaged EE with Eq.~\eqref{eq:fit_function}. 
To show this fact, we concentrate on the case $\gamma=0.1$ and $h=0.5$. In Fig.~\ref{long:fig}(a) we show numerical data
for different power-law exponents $\alpha$ (circles) and the corresponding fit (lines),
that nicely reproduces all the curves.  
In Fig.~\ref{long:fig}(b) we show the parameter $b$ vs $\alpha$. The shaded areas locate the two crossover regions,
respectively at $0.5 \lesssim \alpha^\star_1 \lesssim 1$ and at $\alpha^\star_2 \approx 3.2$,
discussed in Ref.~\cite{Russomanno2023_longrange}.
We expect a volume-law behavior for $\alpha < \alpha^\star_1$ and, in fact, we find $b \approx 0$. 
In contrast, for $\alpha > \alpha^\star_2$, we expect an area-law regime and, in fact, we get $b > 1$. 
Finally, {in the intermediate region between area- and volume-law regimes where a subvolume increase in $L$
  was observed, we consistently find an} exponent $b \approx 0.8$, which is roughly constant in all the region.
Regarding the last observation, we point out that in Ref.~\cite{Russomanno2023_longrange} the observed
subvolume growth was faster than the logarithmic one. Moreover, no analytical function to describe the behavior
of the entanglement with the system size was proposed there, while the fitting function Eq.~\eqref{eq:fit_function}
we suggest here well describes this behavior.
Even for this model, qualitatively analogous considerations apply for other values of $\gamma$.

In Fig.~\ref{long:fig}(c) we plot the length scale $L_0$ given in Eq.~\eqref{l0:eqn} versus $\alpha$.
In the range where we can evaluate it, we find it to be smaller than the maximum of the range where the fit
is applied [$L_{\rm max} = 192$, see Fig.~\ref{long:fig}(a)],
marking the reliability of the fit. For $\alpha\leq 1$ it diverges.
This is consistent with the fact that, in this range of $\alpha$, the entanglement grows as a volume-law,
thus the length scale $L_0$ separating the small-size range of linear increase from the large-size range of power-law increase in Eq.~\eqref{eq:fit_function}
entropy diverges. Notice that when $\alpha=1$ one has $b>0$ and $L_0$ is not divergent but finite, although much larger than $L_{\rm max}$. (We do not show it because it is of order $10^{10}$.) This value of $\alpha$ is probably the transition point between volume-law and power-law regimes; Anyway numerics confirms that here the steady-state entanglement entropy behaves linearly on the whole range where the fit is applied, being $L_0\gg L_{\rm max}$.

\section{Nonintegrable models}
\label{models_ni:sec}

Let us now switch to two paradigmatic nonintegrable models, namely, the staggered $t$-$V$ chain and the SYK model.
In both cases, we are forced to resort to exact diagonalization methods in the full many-body Hilbert space,
therefore our numerics cannot go beyond system sizes $L \sim 20$, preventing us from reliably fitting
the data at various values of $L$ with the function in Eq.~\eqref{eq:fit_function}.
Nonetheless, in what follows we show that, for fixed $L$, the asymptotic bipartite EE as a function of
the system-bath coupling $\gamma$ can be fitted reasonably well by a generalized Lorentzian function
\begin{equation}
  \tilde f(\gamma) = \frac{K}{1 + Q \, \gamma^\beta}\,, \qquad (K,\, Q,\, \beta \geq 0).
  \label{eq:fit_function_nonint}
\end{equation}
The usual Lorentzian function is recovered for $\beta = 2$~\cite{disclaimer}.

In Sec.~\ref{XXZ:sec} we describe how $\overline{S}_\ell$ versus $\gamma$ can be fitted
by the function in Eq.~\eqref{eq:fit_function_nonint} for the $t$-$V$ model with onsite dephasing,
while in Sec.~\ref{SYK:sec} we do the same for the SYK model.
In Sec.~\ref{nonintegrable:sec}, we discuss how the parameters $K,\, Q,\, \beta$ depend on the size $L$.
This analysis shows that, in both cases, the dependence of $\overline{S}_\ell$ on $L$ is of the same form
as in Eq.~\eqref{eq:fit_function}; this finding provides us with the rationale for fitting $\overline{S}_\ell$
versus $\gamma$ with a non intuitive function as Eq.~\eqref{eq:fit_function_nonint}.
As a last stage to understand nonintegrable models, in Sec.~\ref{inverse:sec} we focus
on their localization properties, by checking for the scaling of the IPR with the dimension
of the Hilbert space. We find an anomalous delocalization behavior, which is apparently not related
to the behavior of the EE.

\subsection{Staggered $t$-$V$ model with onsite dephasing}
\label{XXZ:sec}

We consider a tight-binding chain with onsite dephasing, described by
\begin{subequations}
  \label{U1_nonint:eqn}
  \begin{align}
    \hat H_{t\mbox{-}V} = \sum_{j=1}^L \Big[ & -\frac{t}{2} \big( \opcdag{j}\opc{j+1} + {\rm h.c.} \big) + W (-1)^j \hat{n}_j \nonumber \\
      &+ V \big( \hat n_j - \tfrac12 \big) \big( \hat n_{j+1} - \tfrac12 \big ) \Big],
    \label{U1_nonint:Ham}\\
    \hat{m}_j &= \hat n_j , \qquad \text{for} \quad j = 1,\,\ldots,\,L\,,
  \end{align}
\end{subequations}
where $t$ has the same meaning of $J$ in Eq.~\eqref{U1_int_Ham} (here we use a different notation
for historical reasons), $W$ denotes the staggered chemical potential, and $V$ the nearest-neighbor
particle interaction strength. The dissipation is the same as in Eq.~\eqref{U1_int:eqn},
and the Hamiltonian Eq.~\eqref{U1_nonint:Ham} reduces to Eq.~\eqref{U1_int_Ham}, when $V=W=0$.
Note that the presence of a quartic term ($V \neq 0$) as in $\hat H^{(4)}$ prevents this Hamiltonian
from being diagonalized with the techniques discussed in Sec.~\ref{models:sec}. In fact, this model is nonintegrable.

As for the integrable tight-binding chain of Eq.~\eqref{U1_int:eqn}, this model exhibits $U(1)$ symmetry,
thus the dynamics conserves the total number $N$ of fermions.
This observation allows us to restrict the dynamics to the sector of the Hilbert space referring to $N$
fixed by the initial condition. In our case we initialize with the Néel state Eq.~\eqref{init:eqn},
that takes into account the presence of $N=L/2$ fermions, hence we can restrict
to the so called half-filling sector, whose Hilbert space dimension is $\mathcal{N}_L=\binom{L}{L/2}$. 
We approach this problem numerically, using the Krylov algorithm implemented in the Expokit package~\cite{Sidje_Expokit},
which allows us to reach sizes up to $L = 20$. This model has been considered in Ref.~\cite{xing2023interactions},
where evidence of both logarithmic and volume-law scaling of the asymptotic averaged EE has been found.

\begin{figure}[!t]
  \includegraphics[width=8cm]{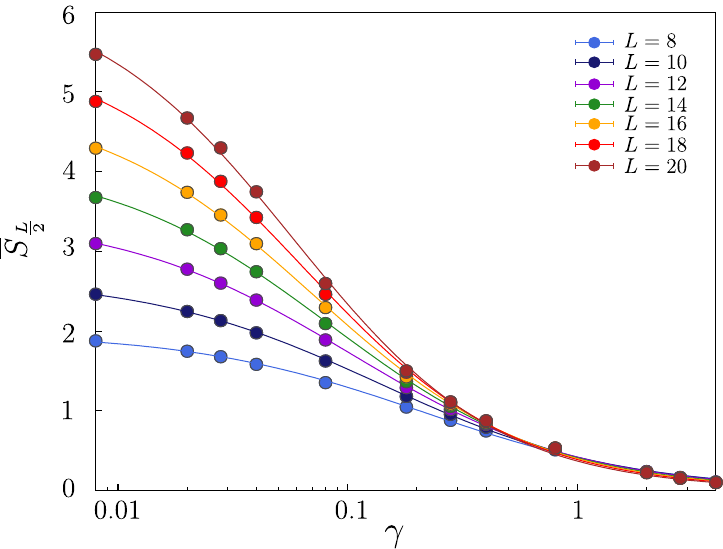}
  \caption{The EE for the model in Eqs.~\eqref{U1_nonint:eqn}.
    Some examples of $\overline{S}_{L/2}$ versus $\gamma$ (circles), for various system sizes
    up to $L=20$ (see legend), together with the corresponding fits with the generalized
    Lorentzian function in Eq.~\eqref{eq:fit_function_nonint} (continuous lines).
    We simulate the time evolution until $t_f = 2 \times 10^3$ with a step $\delta t = 0.01$,
    while we set $W = V = 1$.}
  \label{fig:t-V}
\end{figure}

Figure~\ref{fig:t-V} displays our numerical results for the asymptotic averaged EE $\overline{S}_{L/2}$
versus the measurement strength $\gamma$ (circles) and the corresponding fit obtained
with Eq.~\eqref{eq:fit_function_nonint} (continuous lines).
We can see that the latter performs well over a range of $\gamma \in [8 \times 10^{-3}, 4]$
corresponding to more than two orders of magnitude.

\subsection{SYK model with onsite dephasing}
\label{SYK:sec}

The SYK Hamiltonian is a fermionic long-range interacting lattice model, being characterized by random
four-particle interactions~\cite{PhysRevLett.70.3339,Rosenhaus_2019}.
Adding dissipation in the form of local dephasing, as in Eq.~\eqref{U1_int:eqn}, the model can be written as
\begin{subequations}
  \label{eq:SYK_model}
  \begin{align}
    \hat H_{\rm SYK} & = \frac{1}{\sqrt{L^3}} \sum_{i,j,k,l=1}^L J_{ij,kl} \, \hat c^\dagger_i \, \hat c^\dagger_j \, \hat c_k \, \hat c_l, \label{eq:SYK} \\
    \hat{m}_j & = \hat n_j , \qquad \text{for} \quad j = 1,\,\ldots,\,L\,,
  \end{align}
\end{subequations}
where the couplings $J_{ij,kl}$ are independent Gaussian distributed complex variables, with zero average
$\langle\!\langle J_{ij,kl} \rangle\!\rangle = 0$ and variance
$\langle\!\langle |J_{ij,kl}|^2 \rangle\!\rangle = J^2, \; (J \in \mathbb{R})$. 
The $L^{-3/2}$ prefactor in front of the interaction strength guarantees that the system bandwidth is of the order of $L$,
in the thermodynamic limit $L \to \infty$, such that extensivity of thermodynamic quantities as the energy
is preserved~\cite{Gu_2020,PhysRevB.94.035135,PhysRevB.95.155131}.
Analogously as for the dissipative tight binding chain Eq.~\eqref{U1_int:eqn} and the dissipative $t$-$V$ staggered
model Eq.~\eqref{U1_nonint:eqn},
this model conserves the total number of fermions, thus having a $U(1)$ symmetry. Again, by initializing the system
in the N\'eel state Eq.~\eqref{init:eqn}, we can restrict to the half-filling sector with $L/2$ fermions,
and numerically study the dynamics using the same Krylov algorithm as before.

\begin{figure}[!t]
  \includegraphics[width=8cm]{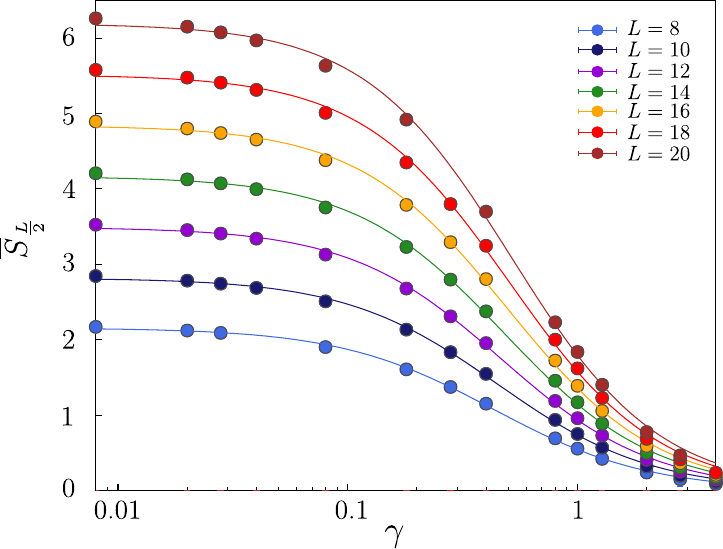}
  \caption{The EE for the model in Eqs.~\eqref{eq:SYK_model}.
    We show some examples of $\overline{S}_{L/2}$ versus $\gamma$ (circles),
    for various sizes up to $L=20$ (see legend), together with the corresponding fits
    with Eq.~\eqref{eq:fit_function_nonint} (continuous lines).
    We simulate the time evolution until $t_f = 3.4 \times 10^2$, with a step $\delta t = 0.01$,
    and fix $J=1$.}
  \label{fig:SYK}
\end{figure}

The SYK model is elusive to perturbative treatments at any energy scale, lying far outside the quasiparticle paradigm.
In fact, it is known to be a paradigm for quantum chaos, displaying fast scrambling~\cite{PhysRevD.94.106002, Roberts_2018},
a nonzero entropy density at vanishing temperature~\cite{PhysRevB.94.035135}, and a bipartite EE
{that scales linearly with $L$} for all the eigenstates (even for the ground state)~\cite{Balents_2018, PhysRevD.100.041901}.
In the context of entanglement transitions, a related model of Brownian SYK chains subject to continuous monitoring
has been considered in Ref.~\cite{PhysRevLett.127.140601}.

In Fig.~\ref{fig:SYK} we show our numerical results for the averaged asymptotic EE
versus the measurement strength (circles) and the fit obtained with Eq.~\eqref{eq:fit_function_nonint}
(continuous lines), displaying again a good agreement between the two, over the same range of
$\gamma$ values as in the $t$-$V$ model.

\subsection{Discussion}
\label{nonintegrable:sec}

\begin{figure}[!t]
  \includegraphics[width=8cm]{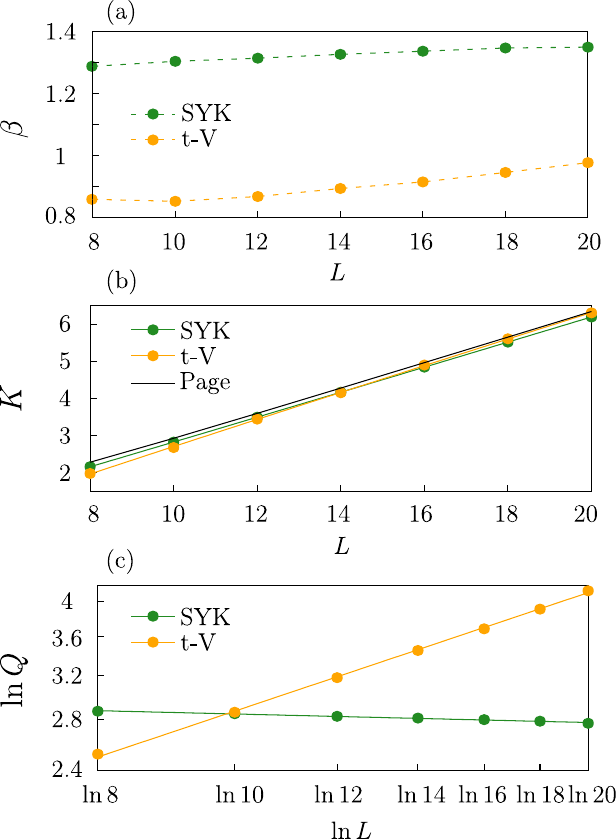}
  \caption{Parameters obtained from the fit with Eq.~\eqref{eq:fit_function_nonint} and plotted against the system size:
    $\beta$ vs $L$ (a), $K$ vs $L$ (b), and $\ln Q$ vs $\ln L$ (c).
    The Page value [black line in (b)] for $S_{L/2}$ is averaged over $N_{\rm r}=48$ realizations
    of a fully random state as in Eq.~\eqref{eq:Randomstate}.
    Panel (c) is in log-log scale.}
  \label{fig:fitting_parameter}
\end{figure}

The results showed in the two previous subsections cannot help too much in determining the asymptotic properties
of the EE. However, some information can be deduced by looking at the fitting parameters. 
In Fig.~\ref{fig:fitting_parameter}, we show the behaviors of $\beta$ vs $L$ (a),
$K$ vs $L$ (b), and of $\ln Q$ vs $\ln L$ (c), for both the $t$-$V$ (orange)
and the SYK (green) models.

Although the reduced sizes we are able to handle are too small for providing a precise statement,
the exponent $\beta$ versus $L$ seems to approach an asymptotic constant value for the SYK model,
while in the $t$-$V$ chain it seems to steadily increase to eventually approach
a linear behavior with increasing size.
On the other hand, it is evident that the parameter $K$ grows almost linearly with $L$,
for both models. More specifically, by fitting the data of Fig.~\ref{fig:fitting_parameter}(b) as
\begin{equation}
  K \sim m L^x + k,
  \label{eq:Kfit}
\end{equation}
we find $x = 1.023 \pm 0.008$ for the $t$-$V$ model, while $x = 0.955\pm 0.039$ for the SYK model. 
Therefore $K$, corresponding to the value of the EE in the $\gamma \to 0$ limit, increases linearly
with the system size $L$. For comparison, the black line also reports the value of $S_{L/2}$
for a fully random state of the form
\begin{equation}
  \ket{\psi} = \frac{1}{\sqrt{\mathcal{N}_L}} \sum_{\{n_j\}} \nep^{-i\varphi_{\{n_j\}}} \ket{\{n_j\}} \,,
  \label{eq:Randomstate}
\end{equation}
where $\ket{\{n_j\}}$ are the simultaneous eigenstates of the operators $\hat{n}_j$ and $\varphi_{\{n_j\}}$
are random phases uniformly distributed in $[0,2\pi]$. This has been worked out some time ago
by Page~\cite{PhysRevLett.71.1291}.
We find that $K$ closely follows the value predicted by Page, suggesting a thermal behavior
of the half-system reduced density matrix in the limit $\gamma\to 0$,
in agreement with previous results on systems obeying eigenstate
thermalization~\cite{Yu_2016, Luitz_2016, PhysRevB.101.064302, PhysRevB.102.144302,Singh_2016}.

Let us now comment on the behavior of $Q$. In fact, as clearly emerging from
Fig.~\ref{fig:fitting_parameter}(c), it behaves quite differently for the two models.
The double-logarithmic plot tells us that this is consistent with $Q \propto L^y$,
a scaling consistent with a behavior described by Eq.~\eqref{eq:fit_function}.
Although the achievable sizes are too small for a large-$L$ extrapolation,
we may obtain an estimate of the exponent $y$ by applying a linear fit to
\begin{equation}
  \ln Q \sim y \ln L + q\,,
\end{equation}
finding
\begin{subequations}
  \begin{eqnarray}
    y & = & 1.57   \pm 0.04  \qquad \qquad (t\mbox{-}V \mbox{ model}) \,, \label{eq:y_tV} \\
    y & = & -0.207 \pm 0.002 \qquad (\mbox{SYK model}) \,. \label{eq:y_SYK}
  \end{eqnarray}
\end{subequations}

Substituting all these findings in Eq.~\eqref{eq:fit_function_nonint},
we can recast the asymptotic EE in the form
\begin{equation}
  \overline{S}_{L/2} \sim \frac{m L^x + k}{1+\gamma^\beta L^y e^q}\,,
\end{equation}
where we used the fact that $x \approx 1$ [Fig.~\ref{fig:fitting_parameter}(b)].
Extrapolating to large $L$, we recover the same dependence on $L$ as in Eq.~\eqref{eq:fit_largeL}:
\begin{equation}
  \overline{S}_{L/2} \sim \frac{\tilde{A}}{\gamma^\beta \tilde{C}} L^{1-b}\,,
\end{equation}
with $\tilde{A} = m$, $\tilde{C} = e^q$, and $b = y$.
Combining this result with those obtained by fitting $Q$, we observe that our procedure predicts
different EE scalings for the two nonintegrable models, in the thermodynamic limit.

\begin{figure}[!t]
  \includegraphics[width=80mm]{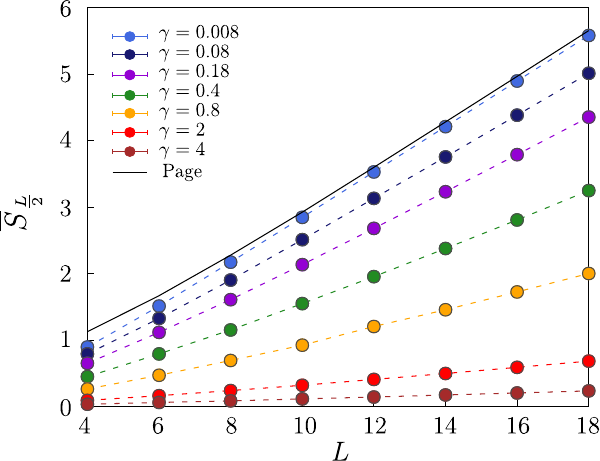}
  \caption{The behavior of $\overline{S}_{L/2}$ versus $L$ for the SYK model.
    Notice the linear increase with the size, after a possible small-size superlinear transient
    for the smaller values of $\gamma$. For comparison, we also report the Page value (black value)
    corresponding to the average over $N_{\rm r}$ fully random states.
    The other parameters are the same as in Fig.~\ref{fig:SYK}.}
  \label{pSLS:fig}
\end{figure}

From the one side, for the SYK model we obtain a superlinear scaling $\overline{S}_{L/2} \sim L^{1.207}$
[cf~Eq.~\eqref{eq:y_SYK}].
Of course, a superlinear growth of the EE cannot be possible for arbitrarily large sizes and, in fact,
it is due to finite-size effects. 
To corroborate this statement, in Fig.~\ref{pSLS:fig} we plot the asymptotic EE of the SYK model versus $L$,
for different values of $\gamma$, and compare with the value predicted by Page~\cite{PhysRevLett.71.1291}
for a random state as in Eq.~\eqref{eq:Randomstate} (black line):
after an initial superlinear transient, which can be better appreciated for small values of $\gamma$,
all the curves approach a linear behavior that is below the Page value. 
This result shows that the fully chaotic nature of the SYK model~\cite{PhysRevD.94.106002,Roberts_2018}
(namely, all its eigenstates {show an EE linear in the system size}~\cite{Balents_2018, PhysRevD.100.041901})
is so robust to survive the measurement process and to lead to a linear increase of the steady-state entanglement
with the size, independently of the measurement strength $\gamma$.

From the other side, for the staggered $t$-$V$ model we find a very different behavior.
Since $1-b \approx -0.57$ [cf~Eq.~\eqref{eq:y_tV}], at some point the EE should start decreasing.
This is likely to be ascribed to a finite-size effect: For larger sizes the fit with Eq.~\eqref{eq:fit_function_nonint}
might not work anymore. This is corroborated by the fact that, in this case, $\beta(L)$ increases with the system size
[see Fig.~\eqref{fig:fitting_parameter}(a)] and does not saturate, so the correct form
is $\overline{S}_{L/2} \sim L^{-0.57}/\gamma^{\beta(L)}$.
This means that, for $\gamma <1$, the increase of $\beta$ might compensate the decrease of $L^{-0.57}$
and the area-law behavior might survive only for $\gamma > 1$. Therefore, our results suggest the presence
of an entanglement crossover from an area-law behavior, for $\gamma \gtrsim 1$, to a regime characterized by
some kind of entanglement increase, for $\gamma \lesssim 1$.
Unfortunately, our numerics does not allow us to make any precise statement on that.

\subsection{Inverse participation ratio and localization properties}
\label{inverse:sec}

Here we consider the inverse participation ratio (IPR), defined as
\begin{equation}\label{IPR:eqn}
  {\rm IPR}(t) = \sum_{\{n_j\}} \big| \braket{\{n_j\}|\psi(t)} \big|^4 \,,
\end{equation}
where $\{n_j\}$ are the ``classical'' configuration states with $n_j$ fermions on the $j$th site,
being simultaneous eigenstates of all the operators $\hat{n}_j$.
The IPR, introduced in Ref.~\cite{Edwards_JPC72}, is a standard measure of delocalization and does not scale
with the dimension of the Hilbert space in the case of perfect localization, while it scales as the inverse
of this dimension in the case of perfect delocalization.
We consider the $t$-$V$ model [Sec.~\ref{XXZ:sec}], its integrable version for $V=0$ (where the quartic terms disappear
and a description as in Sec.~\ref{U1:par} is possible) and the SYK model [Sec.~\ref{SYK:sec}].
All these models conserve the number of fermions, thus the dimension of the Hilbert subspace involved
in the dynamics is $\mathcal{N}_L = \binom{L}{L/2}$. We take the logarithm of the IPR in Eq.~\eqref{IPR:eqn}
and consider its average $\overline{\ln ({\rm IPR})}$ over the quantum trajectories and the time. 
   
\begin{figure}[!t]
  \includegraphics[width=80mm]{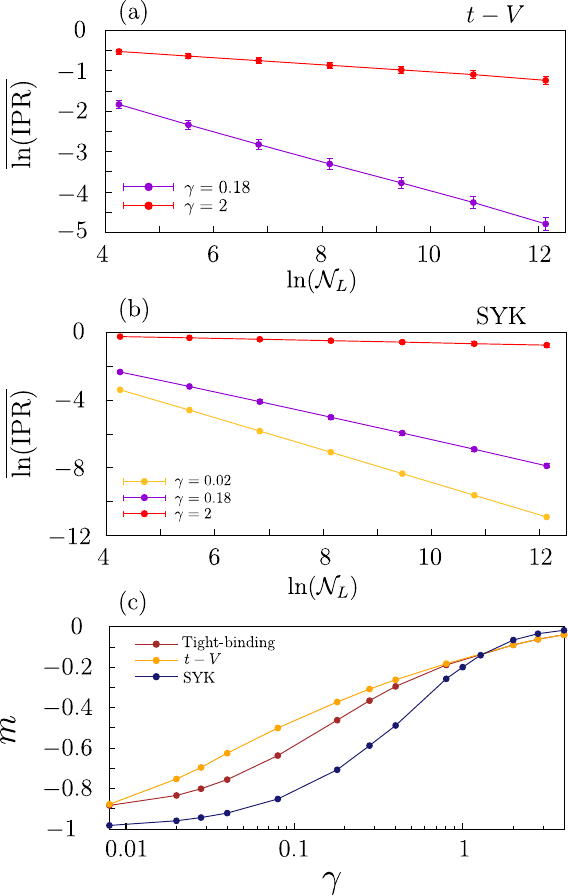}
  \caption{The averaged logarithm of the IPR versus the logarithm of the relevant Hilbert subspace size,
    for (a) the staggered $t$-$V$ chain [Eq.~\eqref{U1_nonint:Ham}, with $W=V=1$]
    and (b) the SYK model [Eq.~\eqref{eq:SYK_model}].
    The various curves for different values of $\gamma$ display a linear dependence with some slope $m$.
    (c) The value of $m$ versus $\gamma$ for the SYK model, the staggered $t$-$V$ chain, and the integrable
    tight-binding chain with a staggered potential [Eq.~\eqref{U1_nonint:Ham}, with $W=1$ and $V=0$].
    We set $t_f = 10^4$ for the staggered chains and $t_f = 1.5 \times 10^3$ for the SYK model, with
    a time step $\delta t = 0.01$.}
  \label{IPR:fig}
\end{figure}

As shown in Fig.~\ref{IPR:fig}(a) for the $t$-$V$ model and in Fig.~\ref{IPR:fig}(b) for the SYK model,
the quantity $\overline{\ln({\rm IPR})}$ behaves always linearly with $\ln (\mathcal{N}_L)$.
We have also analyzed the slope $m$ of this linear dependence versus $\gamma$ [see Fig.~\ref{IPR:fig}(c)] and,
even for this, the behavior is qualitatively the same for all the three cases.
From the one side, $m$ changes smoothly with $\gamma$, independently of the integrability properties and
of the behavior of the EE. From the other side, we always get a value $-1 < m < 0$, meaning that 
the system is neither perfectly delocalized nor perfectly localized. 
   
In summary, in the considered range of $\gamma$, the models we tested are not localized (as shown
in Ref.~\cite{PhysRevLett.127.140601} for the tight-binding case),
as they always display an anomalous delocalization akin to a multifractal
behavior~\cite{PhysRevB.66.033109, PhysRevLett.123.180601, gda_EPJB, chahine2023entanglement}.
We have thus found that, for these monitored systems, localization and delocalization properties
seem to have no relation with the entanglement behavior, although the latter may behave very differently.

\section{Fermionic ladder model}
\label{Sec:Giuliano_model}

\begin{figure}[!t]
  \includegraphics[width=0.8\columnwidth]{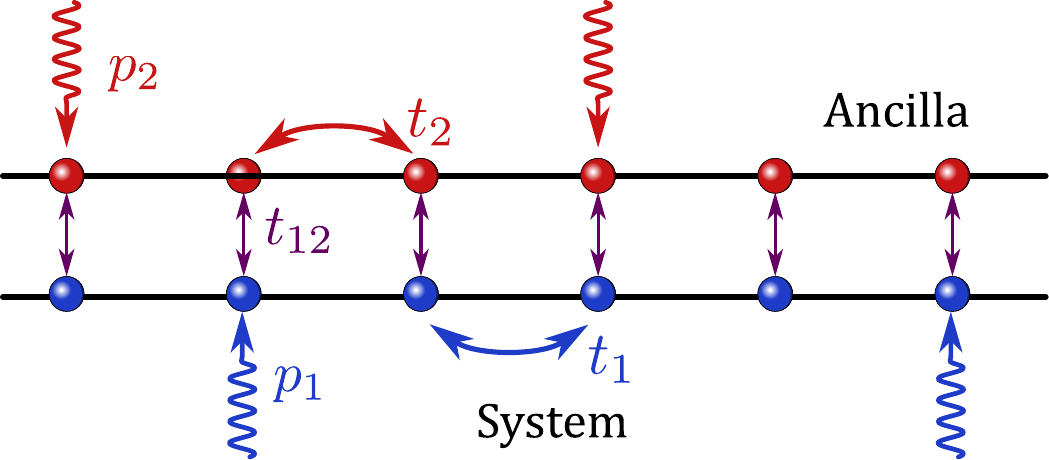}
  \caption{Sketch of the {noninteracting} fermionic ladder model in Eq.~\eqref{Eq:modelH}.
    The blue and red spheres indicate the two chains of fermions representing the System and the Ancilla, respectively.
    Fermions can hop between neighboring sites within the System ($t_1$), the Ancilla ($t_2$), and between
    the System and the Ancilla ($t_{12}$). Wavy lines represent noise acting on the System and the Ancilla.
    After tracing out the Ancilla and partitioning the System into two parts $A$ and $B$,
    we study the entanglement between them.}
  \label{fig:ladder}
\end{figure}

Finally, we test our fitting function on a slightly different model, which has been introduced
and discussed in Refs.~\cite{Tsitsishvili_2024, muzzi2024}. Namely, we consider a {noninteracting} system of two coupled
fermionic chains, each of them with $L$ sites, interacting via local hopping terms,
as shown in Fig.~\ref{fig:ladder}. The quadratic Hamiltonian is given by  
\begin{equation}
  \label{Eq:modelH}
  \hat H_{\rm lad} = \sum_{j,\sigma} t_{\sigma} \big( \hat{c}^{\dagger}_{j,\sigma} \hat{c}_{j+1,\sigma} + \text{h.c.} \big)
  + t_{12} \sum_{j} \big( \hat{c}^{\dagger}_{j,1} \hat{c}_{j,2} + \text{h.c.} \big) \,,
\end{equation}
where $\hat{c}_{j,\sigma}^{(\dagger)}$ are fermionic annihilation (creation) operators on the $j$th site
($j=1,\ldots,L$) of the $\sigma$th chain ($\sigma=1,2$).
The hopping amplitudes within the two chains are $t_1$ and $t_2$, while $t_{12}$ is the interchain hopping amplitude. 
Each chain is subject to periodic boundary conditions,
$\hat{c}^{(\dagger)}_{L+1,\sigma} \equiv \hat{c}^{(\dagger)}_{1,\sigma}$.
Chain 1 is referred to as \textit{the System}, while chain 2 acts as \textit{the Ancilla};
the global system is referred to as \textit{the Ladder}, due to the geometry of the coupling.
The noise is modeled via random projective measurements of the particle number,
$\hat n_{j,\sigma} = \hat{c}^\dagger_{j,\sigma} \hat{c}_{j,\sigma}$, with measurement probabilities $p_1$ and $p_2$
for the System and the Ancilla, respectively.

\begin{figure*}[!t]
  \includegraphics[width=\textwidth]{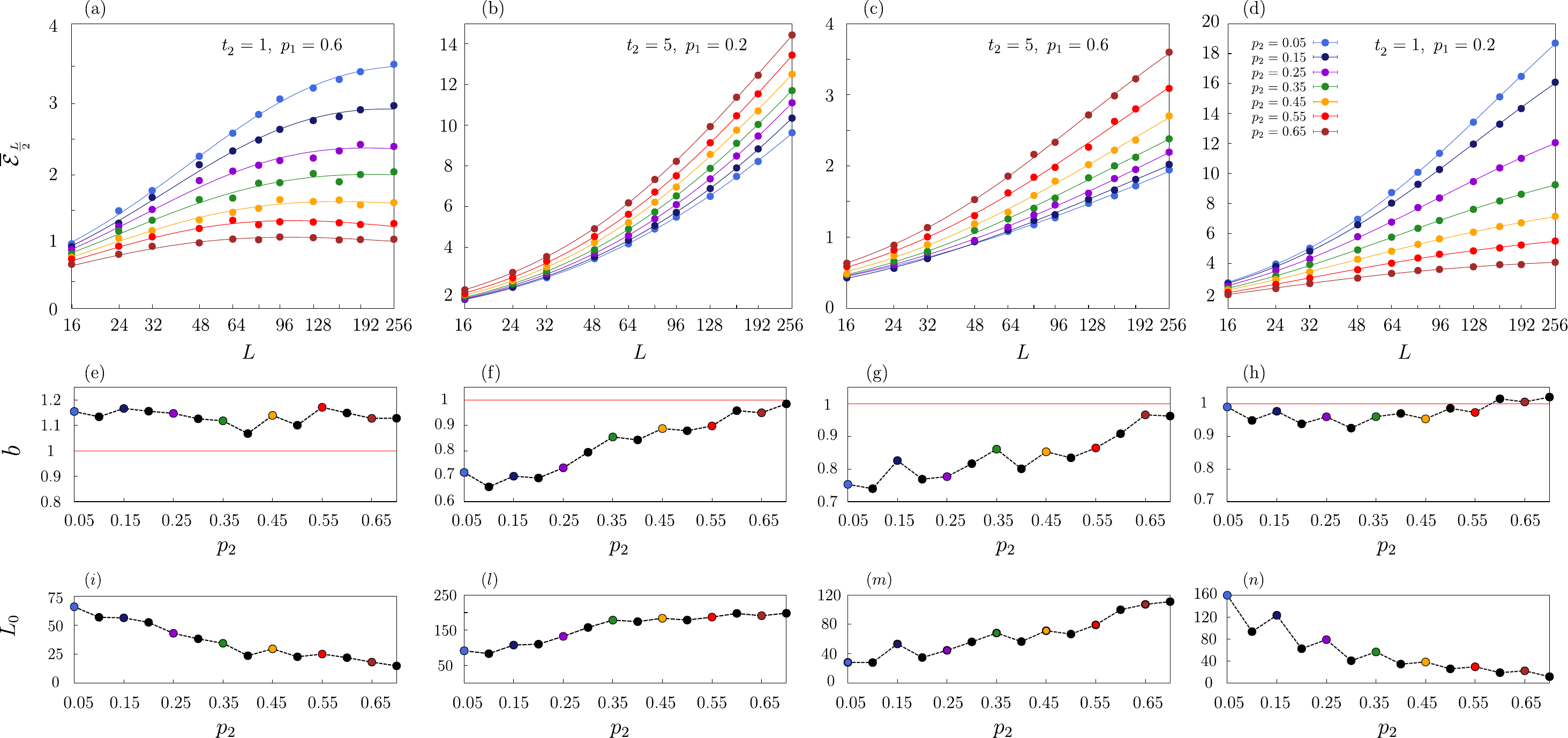}
  \caption{(a), (b), (c), (d): The FLN $\overline{\mathcal{E}}_{L/2}$ versus $L$,
    for different values of $t_2$, $p_1$, and $p_2$. Circles denote the numerical data, while lines correspond
    to the fitting function.
    (e), (f), (g), (h): The fit parameter $b$ vs $p_2$, for the values of $p_1$ and $t_2$ of the corresponding panel above.
    (i), (j), (k), (l): The length scale $L_0$ given in Eq.~\eqref{l0:eqn} vs $p_2$, for the values of $p_1$ and $t_2$ of the corresponding panel above.
    The other parameters used are $t_1=1$ and $t_2=\pi/2$.}
  \label{fig:negativity}
\end{figure*}

In contrast to what described in the previous sections, here the Ladder undergoes a stroboscopic (and not continuous)
evolution, during which the periodic dynamics consists of alternating unitary evolutions and projective
measurements~\cite{Sierant2021, Chiriaco2023}. 
The global system, prepared in a random product state at half-filling, evolves under $\hat H_{\rm lad}$
for a time $\tau_u$ and is then subject to instantaneous local projective measurements~\cite{Tsitsishvili_2024, muzzi2024}.
The cycle repeats $N_{st}$ times until $\tau_{st}=N_{st}\tau_u$, at which a steady state
is reached (for details on the protocol see Appendix~\ref{App:Evolution}).
The final state of the Ladder is pure $\rho = \ket{\Psi(\tau_{st})}\bra{\Psi(\tau_{st})}$,
while the density matrix reduced to the System, $\rho_1=\text{Tr}_2 \rho$, obtained by tracing out
the Ancilla degrees of freedoms, is generally represented by a mixed state.

We are interested in the entanglement between two halves of the System chain, which can be quantified through the FLN,
an entanglement monotone that, contrary to the EE~\cite{Plenio1998a, Donald2002, Plenio2006introduction,RevModPhys.81.865},
is a suitable entanglement measure for mixed states.  
This is defined as
\begin{equation}
  \mathcal{E}_{\ell} = \ln \, \text{Tr} \big| \rho_1^{R_A} \big| \,,
\end{equation}
where $\rho_1^{R_A}$ is the partial time-reversal transformation of the reduced density matrix $\rho_1$,
operated with respect to the partition $A$ whose length is chosen to be
$\ell = L/2$~\cite{Shapourian2017, Shapourian_2019, Turkeshi22:NegativityFF, Ruggiero2019:NegativityFF, muzzi2024}.
We note that the negativity is usually calculated (for bosons) by looking at the spectrum of the partial transpose
of the density matrix (instead of the partial time-reversal). However, the partial transpose does not preserve
the Gaussianity of the state~\cite{eisler2015partial}, so that the negativity of Gaussian fermions cannot be calculated
from the correlation matrix. However the partial time-reversal transformation~\cite{Shapourian2017} preserves Gaussianity,
meaning that the FLN can be obtained from the correlation matrix (see Appendix~\ref{App:Evolution}).

We look at the steady-state trajectory-averaged negativity, by averaging over $N_{\rm r}$ trajectory realizations
and also over the last $m=5$ time steps after $\tau_{st}$, in order to smooth out fluctuations.
Similarly to Eq.~\eqref{selle:eqn}, we have
\begin{equation}
   \overline{\mathcal{E}}_{L/2} = \frac{1}{m} \sum^m_{s=1} \, \overline{\mathcal{E}_{L/2}(\tau_{st}+s\tau_u)} \,.
\end{equation}
The dynamics induced by the Hamiltonian in Eq.~\eqref{Eq:modelH} is Gaussian preserving.
As detailed in Appendix~\ref{App:Evolution}, this allows us to extract the FLN from the two-point
correlation function~\cite{Shapourian2017, Shapourian_2019, Turkeshi22:NegativityFF, Ruggiero2019:NegativityFF, muzzi2024} 
\begin{equation}
  \mathcal{D}_{ij,\sigma \sigma'}(\tau)= \braket{\Psi(\tau) | \hat{c}^\dagger_{i,\sigma}\hat{c}_{j,\sigma'} | \Psi(\tau)} \,,
  \label{eq:Corr_Fermilad}
\end{equation}
thus allowing for numerics up to large system sizes.
The showed results are obtained for $\tau_{st}=250$ and $N_{r}=150$, to ensure convergence.

This model was studied in Ref.~\cite{Tsitsishvili_2024} and more extensively in Ref.~\cite{muzzi2024},
with the purpose to investigate measurement induced transitions in the presence of nonMarkovian noise. 
A rich phenomenology was observed: for small values of $t_2$, a {crossover} from a logarithmic-
to an area-law scaling of the entanglement is induced either by $p_1$ or by $p_2$.
On the other hand, for large values of $t_2$, the logarithmic behavior persists and is actually enhanced
for strong $p_2$, so that the Ancilla protects the entanglement of the system from noise. 
In particular, the logarithmic scaling is clearly seen at larger system sizes $L\gtrsim80$,
with finite-size corrections at lower $L$. In what follows we fix $t_1 = 1$ and $t_{12}=\pi/2$,
to maximize the coupling between the chains.

In Fig.~\ref{fig:negativity}, we show the data for different values of $p_1$, $p_2$ and $t_2$ and the relative
fitting curves obtained with Eq.~\eqref{eq:fit_function}, noticing that the numerical data are well described
for all the considered parameters. In the top panels we show the FLN vs $L$ for different $t_2$, $p_1$ (different panels)
and $p_2$ (different colors). In the bottom panels we show the relative fitting exponent $b$ versus $p_2$.
In particular, in panel (a) we study the regime of small $t_2$ and large $p_1$, where the FLN grows at small $L$
and saturates to an area law. This behavior is well fitted by Eq.~\eqref{eq:fit_function}, as also showed by
the values of $b$ which are consistently larger than 1, see panel (e).
In panels (b) and (c), in correspondence of large $t_2$, we observe a regime where $b$ is significantly smaller
than one [panels (f) and (g)], corresponding to a regime where the asymptotic FLN scales logarithmically with the system size.
Finally for smaller $t_2$ and small $p_2$ [panel (d)], we distinguish both an area law at large $p_2$
and a logarithmic growth at small $p_2$, a behavior recalling the plots in Fig.~\ref{fitI:fig}(a).
However, differently from Fig.~\ref{fitI:fig} where no transition exits and one has always an area law with $b=1$, in this case the exponent is monotonous with the transition parameter,
marking a difference between the two cases. Moreover, $b$ is always close to one [panel (h)], making it difficult
to locate the exact value of $p_2$ corresponding to the crossover between $b>1$ and $b<1$.
Indeed, while the analysis based on the fit with Eq.~\eqref{eq:fit_function} locates the crossover at $p_2 \approx 0.5$,
a refined analysis proposed in Ref.~\cite{muzzi2024} signals the emergence of the transition
from a logarithmic to an area phase at smaller values $p_2 \approx 0.25$.
In panels (i)-(l) we plot $L_0$, i.e., the length scale defined in Eq.~\eqref{l0:eqn}
and separating the volume-law from the power-law entanglement behavior, versus $p_2$.
We always find a value of $L_0$ smaller than $L_{\rm max}=256$ (the maximum of the range where we apply the fit),
confirming that the fit is reliable also in this case.

\section{Conclusions}
\label{conc:sec}

In summary, we have proposed the function in Eq.~\eqref{eq:fit_function} to describe the behavior of the steady-state
long-time EE in monitored fermionic systems, which interpolates between a linear behavior, at small $L$,
and a power-law behavior, at large $L$.
Up to the sizes one can reach with state-of-the-art numerical techniques ($10^1 \lesssim L \lesssim 10^{3}$),
we are able to recover a correspondence between the parameters of the function and some entanglement scaling laws
already known in literature (from area-law, to logarithmic, subvolume-law, and eventually volume-law behavior).

We have tested our function by fitting, in different integrable and nonintegrable models, the steady-state EE
attained by evolving under a quantum-state-diffusion dynamics. More specifically, in the nonintegrable cases,
we have fitted the steady-state EE versus the coupling $\gamma$ with the environment using a generalized
Lorentzian function, and we have recovered the behavior described by Eq.~\eqref{eq:fit_function}.
In particular, we have chosen three integrable one-dimensional models (namely, the tight binding chain with onsite dephasing,
the Kitaev chain both with onsite dephasing and with long-range dissipators) and on two nonintegrable models
(namely, the staggered $t-V$ chain and the SYK model). 
We have also tested our function in a noninteracting fermionic model {on a two-leg ladder},
finding that it also provides a good description of the scaling of the long-time fermionic logarithmic negativity,
suggesting that our result is a good indicator of the entanglement scaling,
independently of the monotone considered. 
In all the above cases, we have found a good qualitative agreement with the existing knowledge of the entanglement
behavior with the system size. Note that the logarithmic growth with $L$, although not explicitly present
in our formula of Eq.~\eqref{eq:fit_function}, can be glimpsed by a power-law fitting behavior
with an exponent $b \approx 0.8$. On the basis of a purely numerical analysis, one cannot rule out that this might also
be due to a finite-size effect asymptotically providing an area law.
  
Let us stress again that, due to the lack of a proper analytical support, the results presented here
should not be intended as suitable for predicting any otherwise unknown entanglement phases.
However, given the reliability of Eq.~\eqref{eq:fit_function} in capturing the entanglement behavior
for a variety of different models in a fairly wide range of system sizes,
we think that this result may contribute to the development of the theory of entanglement transitions in monitored systems.
We are aware of already existing conformal field theory descriptions of this phenomenon, consistent
with a large-$L$ behavior. 
We think it is however worth investigating whether it would be possible to formulate a theory that can incorporates
the small-size behavior not only as ``corrections''. Moreover in some cases, as for the description of the
intermediate regime in the Kitaev chain with long-range dissipator, the fitting function~\eqref{eq:fit_function}
is likely to perform better than the usual logarithmic scaling guess.  

Characterizing the short-size behavior of the entanglement can also be useful from an experimental point of view.
In fact, if one could find a way to extrapolate information on the entanglement scaling by looking at the behavior
for small sizes, it would be then easier to access any experimental verification with present-day technologies~\cite{G_AI}.

\acknowledgments

We thank M.~Fava, I.~V. Gornyi, A.~D. Mirlin, and M.~Szyniszewsi for fruitful discussions.
A.~R. acknowledges computational resources from MUR, PON “Ricerca e Innovazione 2014-2020”,
under Grant No. PIR01 00011 - (I.Bi.S.Co.).
We acknowledge support from the Italian MUR through the following projects: PRIN 2017 No. 2017E44HRF and PNRR MUR project PE0000023-NQSTI.
G.C. is supported by ICSC – Centro Nazionale di Ricerca in High-Performance Computing,
Big Data and Quantum Computing under project E63C22001000006.

\appendix


\section{Fit stability}
\label{App:stability}

Here we provide some arguments regarding the stability of the fit proposed in Eq.~\eqref{eq:fit_function}. 
In particular, we focus on the stability of the parameter $b$, by fitting the same data of Fig.~\ref{fig:ising}
in a range $[L_\text{min}, L_\text{max}]$, with varying $L_\text{min}$ and $L_\text{max}$.

\begin{figure}
  \includegraphics[width=\columnwidth]{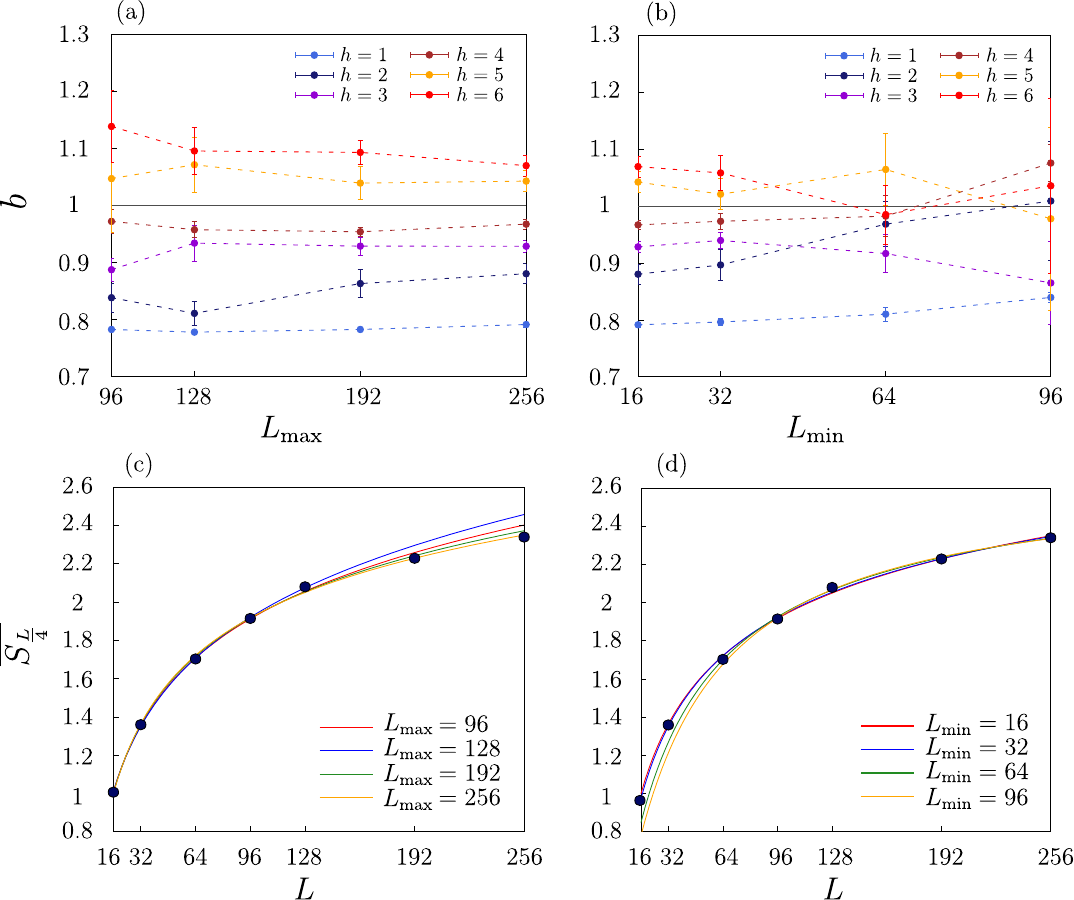}
  \caption{Top panels: The exponent $b$ as a function of $L_\text{max}$ (a) and $L_\text{min}$ (b),
    for the same data of Fig.~\ref{fig:ising}. The black line marks the value $b = 1$.
    Bottom panels: The various fitting functions for the numerical data of the EE with $h=2$ (black circles),
    as obtained by changing the values of $L_\text{max}$ (c) and of $L_\text{min}$ (d).
	  }
  \label{fig:fit_stability}
\end{figure}

The results are showed in Fig.~\ref{fig:fit_stability}. In the top panels we report the value of $b$ obtained 
by a fit of the numerical data for the EE $\overline{S_{L/4}}$ in a range of system sizes from $L=16$
to $L=256$, constraining the fit either from $L_\text{min} = 16$ to a varying size $L_\text{max}$ [panel (a)],
or from a varying size $L_\text{min}$ to $L_\text{max} = 256$ [panel (b)].
We notice that the fit parameter remains more stable when small sizes are taken into account.
In fact, as emerging from panel (b), if $L_\text{min}$ is too large, one can also predict
a wrong entanglement behavior (i.e., the fitted value of $b$ can become smaller or larger than one,
thus signaling a change of behavior from subvolume-law to area-law). 
This result suggests that, differently from the logarithmic fit currently employed in the literature,
our procedure is rather sensitive to the behavior for the EE at smaller system sizes,
compared to the one at larger sizes. It is thus important to obtain a good knowledge of the short-size
behavior ($L \leq 100$), which is more easily accessible by numerical approaches.

To test the quality of our findings, in panels (c) and (d) we have plotted the best fit function for
the data with $h=2$. The different curves have been obtained either by varying $L_\text{max}$ and
fixing $L_\text{min}=16$ [panel (c)], or by varying $L_\text{min}$ and fixing $L_\text{max}$ [panel (d)].

\begin{figure*}[!t]
    \includegraphics[width=\textwidth]{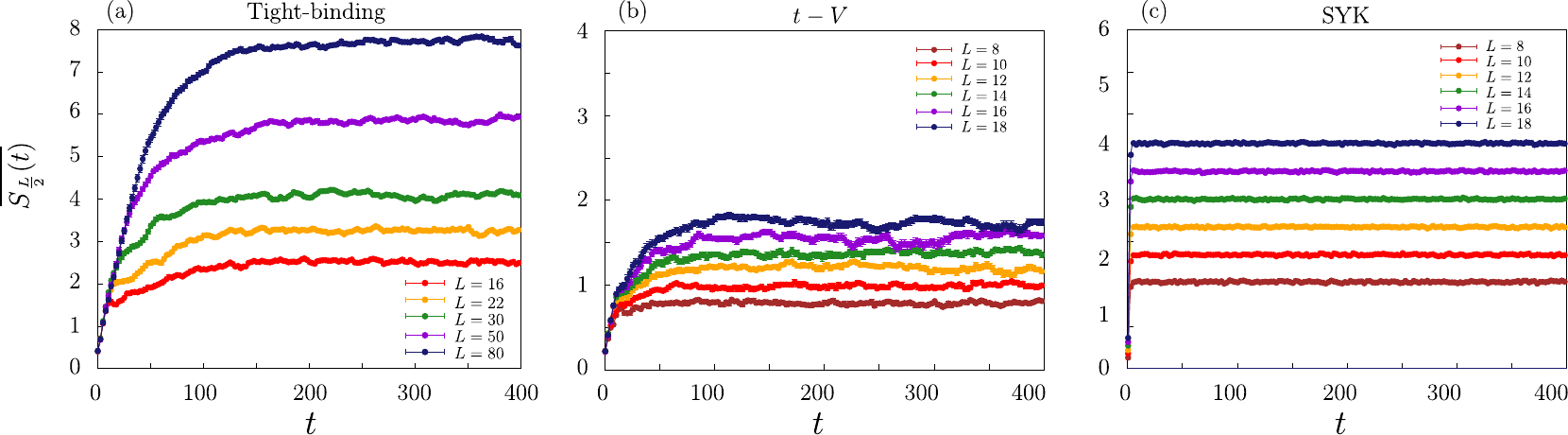}
    \caption{The behavior of $\overline{S_{L/2}(t)}$ versus $t$ for (a) the staggered tight-binding chain
      [Eq.~\eqref{U1_nonint:Ham}, with $W=1$ and $V=0$], (b) the staggered $t$-$V$ chain [Eq.~\eqref{U1_nonint:Ham},
        with $W=V=1$], and (c) the SYK model [Eq.~\eqref{eq:SYK_model}].
      We choose $\gamma = 0.04$ and report data for different system sizes (see legend).
      The time step has been fixed as $\delta t =10^{-2}$ and, in all cases, is expressed in units of $J=1$.
      We take $N_{\rm r} = 48$ in panels (a) and (c), and $N_{\rm r}= 64$ in panel (b).}
    \label{fitit:fig}
\end{figure*}

\section{Time traces}
\label{ttraces:sec}

Some examples of the time traces for the trajectory-averaged half-chain EE, $\overline{S_{L/2}(t)}$,
are shown in Fig.~\ref{fitit:fig}. We present results for (a) the integrable staggered tight-binding model,
(b) the nonintegrable staggered $t$-$V$ model, and (c) the SYK model.

We can observe that, for the SYK model, the EE saturates the fastest and displays the smallest fluctuations in time.
Although in the presence of measurements ($\gamma = 0.04$), the SYK model shows a dramatically fast relaxation
to the stationary long-time limit value $\overline{S_{L/2}}$ and is self-averaging.
On the other hand, the mere absence of integrability does not qualitatively change the main features
of the time traces for $\overline{S_{L/2}(t)}$ [compare (a) with (b), where the only difference is to choose
$V=0$ or $V=1$ in Eq.~\eqref{U1_nonint:Ham}, respectively].
We recall that the results presented in the main text are obtained by further averaging such curves
over the time. This double averaging process, over the trajectories and time, allows us 
to smoothen fluctuations and to get rid of them (we checked that the presented results
are stable by further increasing $N_r$).

\section{Fermionic logarithmic negativity in the ladder model}
\label{App:Evolution}

Following Ref.~\cite{muzzi2024}, we first show how the correlation matrix~\eqref{eq:Corr_Fermilad},
for the free-fermion model on a two leg ladder described in Sec.~\ref{Sec:Giuliano_model}, evolves
under the combined action of the unitary dynamics and the measurements.
Our protocol is composed of (i) a unitary dynamics generated by the Hamiltonian in Eq.~\eqref{Eq:modelH},
$\hat U = e^{-i \hat H_{\rm lad} \tau_u}$, and (ii) a sequence of measurements of the fermionic number $\hat{n}_{j,\sigma}$
on each site, with probability $p_\sigma$ (for $\sigma=1,2$).

The state of the ladder after the unitary evolution is given by $\ket{\Psi(\tau_u)}=\hat{U}\ket{\Psi(0)}$. 
Going in the Fourier-Nambu space we can write 
\begin{equation}
  \notag\hat{H}_{\rm lad} = \sum_{k} \hat{\psi}^\dagger_k \, \mathbb{H}_k \, \hat{\psi}_k \,,
  \label{H_k}
\end{equation}
where
\begin{equation}
  \mathbb{H}_k = \begin{pmatrix}
    2t_1\cos k & t_{12}\\
    t_{12} & 2t_2 \cos k
  \end{pmatrix}
\end{equation}
and
\begin{equation}
  \hat{\psi}^{\dagger}_k \equiv (\hat{c}^{\dagger}_{k,1}, \;\; \hat{c}^{\dagger}_{k,2}) \,, \quad
  \hat{c}_{k,\sigma} = \frac{1}{\sqrt{L}}\sum_j e^{-ijk} \hat{c}_{j,\sigma}\,,
\end{equation}
is the Nambu spinor in Fourier space.
In this way, we can factorize the unitary evolution operator as $\hat{U} = \otimes_k \hat{U}_k$,
where $\hat{U}_k = e^{-i \hat{H}_k \tau_u}$ can be written using an explicit analytic expression~\cite{Tsitsishvili_2024, muzzi2024}.

During the unitary part of the evolution, the correlation matrix $\mathcal{D}(\tau)$ can be thus shown
to change according to~\cite{Coppola2022, muzzi2024}
\begin{equation}
  \mathcal{D}(\tau+\tau_u) = \hat{\mathbb{R}}^{\dagger}\mathcal{D}(\tau)\hat{\mathbb{R}} \,,
\end{equation}
with
\begin{equation}
  \hat{\mathbb{R}}_{mn}=\frac1L\sum_ke^{-ik(m-n)}\hat{U}_{k} \,.
\end{equation}
For what concerns the impact of measurements, the operators $\hat n_{l,\mu}$ and $1 - \hat n_{l,\mu}$
are orthogonal projectors, thus the probability to measure $n_{l,\mu}=1$ is given by
$p_{n_{l,\mu}=1}(\tau) = \bra{\Psi(\tau)} \hat n_{l,\mu} \ket{\Psi(\tau)}$, while the probability
to measure $n_{l,\mu}=0$ is given by $p_{n_{l,\mu}=0}(\tau) = 1-p_{n_{l,\mu}=1}(\tau)$.
The effect of the measurements translates into the following update rule
for the correlation matrix $\mathcal{D}_{ij,\sigma\sigma'}(\tau)$~\cite{muzzi2024}:

\begin{enumerate}
\item
  For each site $l$ belonging to chain $\mu$, extract a random number $z_{l,\mu} \in (0,1] $.
  If $z_{l,\mu} \leq p_\mu$, the measurement is performed.
    
\item
  If the measurement must be performed, extract a second random number $q_{l,\mu} \in (0,1]$.
    
\item
  If $q_{l,\mu} \leq p_{n_{l,\mu}=1}(\tau)$, then the operator $\hat{n}_{l,\mu}$ is applied to the state:
  \begin{equation}
    \ket{\Psi(\tau)} \mapsto \frac{ \hat{n}_{l,\mu}\ket{\Psi(\tau)} }{ \| \, \hat{n}_{l,\mu}\ket{\Psi(\tau)} \| } \,,
  \end{equation}
  which, thanks to Wick's theorem, results into
\begin{equation}
    \begin{split}
      \mathcal{D}_{ij,\sigma \sigma'}(\tau) \to & \mathcal{D}_{ij,\sigma \sigma'}(\tau)
      + \delta_{il}\delta_{jl}\delta_{\sigma \mu}\delta_{\sigma' \mu} 
      \\
      & - \frac{\mathcal{D}_{il,\sigma \mu}(\tau) \, \mathcal{D}_{lj,\mu \sigma'}(\tau)}{\mathcal{D}_{ll,\mu\mu}(\tau)}.
    \end{split}
\end{equation}

\item
  If $q_{l,\mu} > p_{n_{l,\mu}=1}(\tau)$, then the operator $1 - \hat{n}_{l,\mu}$ is applied to the state:
  \begin{equation}
    \ket{\Psi(\tau)} \mapsto \frac{ (1-\hat{n}_{l,\mu})\ket{\Psi(\tau)} }{ \| (1-\hat{n}_{l,\mu})\ket{\Psi(\tau)} \| } \,,
  \end{equation}
  which results into 
\begin{equation}
    \begin{split}
        \qquad & \mathcal{D}_{ij,\sigma \sigma'}(\tau) \to
        \mathcal{D}_{ij,\sigma \sigma'}(\tau) - \delta_{il}\delta_{jl}\delta_{\sigma \mu}\delta_{\sigma' \mu} 
        \\
        & \;\; + \frac{(\delta_{il,\sigma \mu}-\mathcal{D}_{il,\sigma \mu}(\tau))(\delta_{lj,\mu\sigma'}-\mathcal{D}_{lj,\mu \sigma'}(\tau))}{1-\mathcal{D}_{ll,\mu\mu}(\tau)} \,.
    \end{split}
\end{equation}

\end{enumerate}

The FLN can be obtained through the spectrum of the correlation matrix $\mathcal{D}(\tau)$,
reduced to the degrees of freedom of the system.
In particular $\mathcal{E}= \ln \Tr |\rho^{R_A}_1|= \ln \Tr \sqrt{\rho^{R_A}_1 \big( \rho^{R_A}_1 \big)^\dagger}$,
where $\rho^{R_A}_1$ is the partial time reversal of the reduced density matrix of the system $\rho_1$,
with respect to the subsystem A. Since the partial time reversal transpose preserves the Gaussianity of the state,
then also $\rho^{R_A}_1$ and the product $\rho^{R_A}_1 \big( \rho^{R_A}_1 \big)^\dagger$ are Gaussian,
so that their spectral properties can be calculated from the correlation matrix.

We define $\mathcal{D}_{1,ij}\equiv\mathcal{D}_{ij,11}$ the correlation matrix restricted to the System and introduce 
\begin{equation}
    \Gamma_{1,ij}= 2\mathcal{D}_{1,ij}-\delta_{ij}.
\end{equation}
Given a bipartition of the System into subsystems $A$ and $B$, the matrix $\Gamma_1$ takes the block form
\begin{equation}
    \Gamma_1 = \begin{pmatrix}
        \Gamma_{1,AA} & \Gamma_{1,AB}\\
        \Gamma_{1,BA} & \Gamma_{1,BB}
    \end{pmatrix} .
\end{equation}
We also introduce the correlation matrices 
\begin{equation}
    \Gamma_{\pm} = \begin{pmatrix}
        \Gamma_{1,AA} & \pm i \, \Gamma_{1,AB}\\
        \pm i \, \Gamma_{1,BA} & -\Gamma_{1,BB}
    \end{pmatrix}
\end{equation}
associated with $\rho^{R_A}_1$ and $(\rho^{R_A}_1)^\dagger$.

The FLN is then computed from the eigenvalues $\{\lambda_j\}$ of $\mathcal{D}_1$ and from the eigenvalues
$\{\mu_j\}$ of $\Gamma_\times$, defined as~\cite{Fagotti_2010,EisertNegativity}
\begin{equation}
  \Gamma_\times = \frac{1}{2} \big[ 1-(1+\Gamma_+\Gamma_-)^{-1}(\Gamma_+ + \Gamma_-) \big] \,,
\end{equation}
in particular it holds \cite{Shapourian_2019}
\begin{equation}
  \mathcal{E}_A = \sum^{L}_{j=1} \biggr\{ \ln \big( \sqrt{\mu_j}+\sqrt{1-\mu_j} \big)
  + \frac{1}{2} \ln \big[ (1-\lambda_\alpha)^2+\lambda^2_\alpha \big] \biggr\} .
\end{equation}

%

\end{document}